\documentclass[]{emulateapj}

\usepackage{graphics}
\usepackage{natbib}
\usepackage{ulem}
\usepackage{soul}
\usepackage{amsmath, amsthm, amssymb,color}

\newcommand{\ba}{\begin{eqnarray}}
\newcommand{\ea}{\end{eqnarray}}

\newcommand{\out}{_\mathrm{out}}

\newcommand{\evec}{\mathbf{e}}
\newcommand{\jvec}{{\mathbf{j}}}

\shorttitle{Merging compact-object binaries in non-spherical nuclear star clusters}
\shortauthors{Petrovich \& Antonini}

\begin{document}
\title{Greatly enhanced merger rates of compact-object binaries 
in non-spherical nuclear star clusters}
\author{ Cristobal Petrovich\altaffilmark{1,2} \& Fabio Antonini\altaffilmark{3}}
\altaffiltext{1}{Canadian Institute for Theoretical Astrophysics, University of Toronto, 
60 St George Street, ON M5S 3H8, Canada; cpetrovi@cita.utoronto.ca}
\altaffiltext{2}{Centre for Planetary Sciences, Department of Physical \& 
Environmental Sciences, University of Toronto at Scarborough, Toronto, 
Ontario M1C 1A4, Canada}
\altaffiltext{3}{Center for Interdisciplinary Exploration 
and Research in Astrophysics (CIERA) and department of physics and astronomy,
Northwestern University, USA; fabio.antonini@northwestern.edu}

\begin{abstract}
The Milky Way and a significant fraction 
of galaxies are observed to host a central Massive Black Hole 
(MBH) embedded in a non-spherical 
nuclear star cluster.
We study the secular orbital evolution of 
compact-object binaries in 
these environments  and characterize the excitation of extremely large 
eccentricities that can lead to mergers by gravitational radiation.
We find that the eccentricity excitation occurs
most efficiently when the nodal precession
timescale of the binary's orbit around
the MBH due to the non-spherical cluster becomes comparable 
(within a factor of $\sim10$) to the 
timescale on which the binary is torqued by the MBH due 
to the Lidov-Kozai (LK) mechanism. 
We show that in this regime
the perturbations due to the cluster
increase the fraction of systems that reach extreme eccentricities 
($1-e\sim10^{-4}-10^{-6}$) by a factor of $\sim10-100$ compared to
the idealized case of a spherical cluster,
increasing the merger rates of compact objects by a similar factor.
We identify two main channels
that lead to this extreme eccentricity excitation: (i) chaotic 
diffusion of the eccentricities due to resonance
overlap; (ii) cluster-driven variations of the mutual inclinations
between the binary orbit
and its center-of-mass orbit around the MBH, which can
intensify the LK oscillations.
We estimate that our mechanism can produce black hole-black hole and 
black hole-neutron star binary merger
rates of up to $\approx15\,\rm{Gpc}^{-3}yr^{-1}$ and $\approx0.4\,\rm{Gpc}^{-3}yr^{-1}$,
respectively.
Thus, we propose the {\it cluster-enhanced} Lidov-Kozai 
mechanism as a new channel for the merger of 
compact-object binaries, competing with scenarios that invoke
isolated binary evolution or dynamical formation in 
globular clusters.
\end{abstract}

\section{Introduction}

Most nearby galaxies contain a compact stellar cluster residing at their kinematical and photometric center 
\citep[e.g.,][]{2011MNRAS.413.1875N,2012ApJS..203....5T,2014MNRAS.445.2385D,2014MNRAS.441.3570G,2016MNRAS.457.2122G}.
These nuclear clusters (NCs) have masses in the range 
$\approx 10^5-10^8~ M_\odot$ and half mass radii of only a few parsecs, making them the most 
massive and densest stellar clusters observed in the local Universe \citep{2008MNRAS.389.1924F,2011MNRAS.414.3699M}. 
Most NCs are believed  
 to host a central massive black hole (MBH); this is the case for the Milky Way, the location of
  a $\sim 3\times 10^7M_\odot$ NC
  \citep{2014A&A...566A..47S,2017arXiv170103817S} 
and a $\approx 4\times 10^6M_\odot$ MBH  \citep{2008ApJ...689.1044G,2009ApJ...692.1075G}.
Examples of galaxies hosting both a compact NC and an AGN are known across a wide range of galaxy masses
\citep{2008ApJ...678..116S,2012AdAst2012E..15N},  suggesting that 
NCs and MBHs often co-exist in galaxies.

The dynamics of stars near a MBH in a NC
 is characterized by high velocity dispersion,  extreme stellar densities and short relaxation time
 \citep{2013degn.book.....M}. 
On even shorter timescales stars move around the MBH on nearly Keplerian orbits.
Binaries that reside in this environment can therefore interact with the MBH for many orbits
and perturbations resulting from their interaction with the MBH can accumulate
for a long time before the binary properties are  significantly altered by encounters with other field stars.
In particular, binaries with highly inclined orbits
with respect to their orbit around the MBH can  be strongly affected by 
secular processes, namely by Lidov-Kozai (LK) oscillations, which
periodically change their eccentricities and inclinations 
\citep{1962P&SS....9..719L,1962AJ.....67..591K}. 
At the highest eccentricities during a LK oscillation
the components of the binary can merge \citep{2010ApJ...713...90A,2012ApJ...757...27A}.
This process can lead to a number of potentially important phenomena, including the merger of stellar binaries which can result in the formation of 
massive stars and/or G2-like objects \citep{2011ApJ...731..128A,2015ApJ...799..118P,2016MNRAS.460.3494S, 2017MNRAS.466.3376L};
 and  the formation of X-ray binaries.
Particularly exciting is the possibility  that many, or maybe even most, of the 
merging compact object binaries that are soon to be detected by Advanced LIGO (aLIGO) \citep{2016PhRvL.116f1102A,2016ApJ...833L...1A} 
are formed near MBHs through the LK mechanism 
\citep{2012ApJ...757...27A,2016ApJ...828...77V}.

In the standard theory of hierarchical triple systems, for an eccentric outer orbit, 
the inner binary can reach  high eccentricities and undergo chaotic evolution 
of its orientation for a large range of initial mutual inclinations of inner to outer orbit 
\citep[e.g.,][]{2000ApJ...535..385F,LN11,katz11,li14}.
However, in the environment of a  MBH, extremely high eccentricities 
are difficult to achieve for most binaries 
\citep{2012ApJ...757...27A}. 
One reason for this is that  the
octupole-order terms are effectively vanishing in this case (see also Section \ref{Nprec} below).
Following \citet{naoz13}, we quantify the importance 
of the octupole-level coefficients
using the quantity
\begin{equation}
\epsilon_{\rm oct} = \left( m_1-m_2 \over m_1+m_2\right) 
\left(a_{\rm in}\over a_{\rm out}\right) {e_{\rm out}\over 1-e_{\rm out}^2} 
\label{eq:oct}
\end{equation}
with $m_1$ and $m_2$ the mass of the binary components, $a_{\rm in}$ the 
binary semi-major axis, and $e_{\rm out}$ and $a_{\rm out}$ the eccentricity
and semi-major axis of the orbit of the binary
around the MBH, respectively.
Even for large values of $e_{\rm out}\sim 0.9$,
and for $a_1\sim 1 $ AU
the octupole-order terms are
effective (i.e., $\epsilon_{\rm oct}\gtrsim 10^{-3}$) 
 only  within a  short distance  from the galactic center ($\lesssim 0.01\rm pc$), which  for a Milky Way like galaxy is inhabited only by a tiny  fraction of 
the overall population of stars  bound to the MBH \citep[see Figure 14 in][]{2015A&A...584A...2F}.
\footnote{This conclusion  is in contradiction  with pevious work 
from \citet{2016MNRAS.460.3494S}
who argued for a key contribution of the octupole-order terms. This in spite of their Figure 1 (top panel) which  shows that the octupole dynamics  can only become
relevant inside $\approx 0.0\rm 1pc$, in agreement with our conclusion.
Moreover, \citet{2016MNRAS.460.3494S} 
and most previous work \citep[e.g.,][]{2015ApJ...799..118P}
did not take into account the mass precession of the outer orbit which
further suppresses the contribution  of the octupole-order terms (See section \ref{Nprec}).}

A direct consequence of the negligible  octupole contribution
is that the magnitude of the inner orbit's angular momentum has a lower
limit equal to its constant component, $j_z$, along the total angular momentum. 
Thus, for isotropic initial conditions, the 
probability of achieving the high eccentricities 
required for a gravitational-wave mediated 
merger, $1-e_{\rm in}\lesssim10^{-5}$ (see Eq. [\ref{eq:e_merge}]), 
is only $\lesssim 0.003$.\footnote{Only binaries with inclinations relative to
their orbit around the MBH in the range $[89^\circ.8,90^\circ.2]$
can reach $1-e_{\rm in}\lesssim10^{-5}$.}

From our discussion and consistent with the
population synthesis study by \citet{2012ApJ...757...27A},
 it follows that  in a spherical star cluster, where 
$j_z= \rm const$, the contribution
from the standard LK mechanism 
 to the number of merging binaries in galactic nuclei
 is negligible.
However, there is ample evidence that a substantial fraction of NCs, 
including the one in the Milky Way,
have a non-spherical shape  \citep[e.g.,][]{2014MNRAS.441.3570G,Chat15,2017MNRAS.466.4040F}. 
One important feature of axisymmetric
(and triaxial) systems 
is that the angular momenta (orientation and/or magnitude) 
of stellar orbits are not conserved. 
In particular, the orientation of the stellar orbits around the MBH
 can change gradually with time, implying  that $j_z\ne \rm const$.
Similarly,  close to the MBH angular momenta can also evolve  due to resonant relaxation \citep{1996NewA....1..149R}.
Consequently,  the available fraction of phase space for
achieving high values of $e_{\rm in}$ can potentially be
much larger than that obtained in the standard LK theory.
Thus, the binary merger rate in a real NC 
can be well above that obtained for an
idealized model of a spherical star cluster. 

In this paper, we consider  the evolution of binaries residing in
a non-spherical NC hosting a central MBH
and argue that the potential from the cluster causes precession of the 
binary's orbit in timescales that are comparable 
to the LK timescale of typical binaries (Eqs. [\ref{eq:tau_lk}]-[\ref{eq:nodal}]). 
As a result,  $j_z$ can change dramatically and even lead to chaotic 
behavior so that extreme eccentricities can be attained for a wide range of
orbital configurations.
We quantify the how this mechanism enhances the merger rates of
binaries comprising neutron stars (NSs) and black hole (BHs) 
in galactic centers.

The paper is organized as follows. 
In \S\ref{sec:basics} we discuss our basic set up with the
equations of motion fully expressed in the Appendix.
In \S\ref{sec:timescales} we describe the relevant timescales 
in the problem.
Our results are presented in \S\S\ref{sec:pop_synth} and
\ref{sec:dynamics}, where the former section presents the 
general dynamical behavior and simplified model, while the latter 
section shows the statistical results from a population 
of binaries changing different parameters. We
summarize the results of these two sections in 
\S\ref{sec:summary}.
In \S\ref{sec:BHrate} we discuss the implications
of our results for the merger rates of compact-object binaries.
We summarize our main results in
\S\ref{sec:conclusion}.

\begin{figure}
\center
\includegraphics[width=9cm]{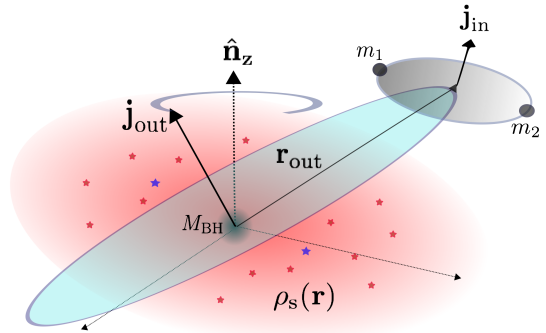}
\caption{Schematic view of the coordinate system and general 
set up. 
The massive black hole of mass $M_{\rm BH}$ sits at the center of 
the coordinate system and the density of the stellar cluster $\rho_s({\bf r})$ is 
axisymmetric along $\hat{\bf n}_z$ 
(we also treat general case of a triaxial cluster). 
The inner orbit corresponds to the binary composed of 
masses $m_1$ and $m_2$
and has angular momentum vector ${\bf j}_{\rm in}$.
The outer orbit corresponds to the binary center of mass 
around $M_{\rm BH}$, which follows
a Keplerian orbit with position ${\bf r}_{\rm out}$
and has an angular momentum vector ${\bf j}_{\rm out}$ that
precesses around $\hat{\bf n}_z$ (i.e., undergoes nodal
precession). }
\label{fig:cluster}
\end{figure}

\section{Basic equations}
\label{sec:basics}

We model the evolution of a stellar binary with component masses
of $m_1$ and $m_2$ with semi-major axes $a_{\rm in}$ orbiting a MBH  of
mass $M_{\rm BH}$ in a Keplerian orbit with semi-major axis 
$a_{\rm out}$.
For the sake of brevity, in some expressions we drop 
the sub-index ``in" when referring to the inner orbit.
Thus,  $e\equiv e_{\rm in}$, $a\equiv a_{\rm in}$,
 $\omega\equiv \omega_{\rm in}$, and  so on.

We express the potential 
per unit mass from the MBH of mass $M_{\rm BH}$ and 
stellar cluster potential that governs the dynamics of a binary 
with masses $m_1,m_2\ll M_{\rm BH}$ as
\ba 
{\Phi}({\bf r})&=&-\frac{GM_{\rm BH}}{r}+\frac{4\pi G}{(3-\gamma)(2-\gamma)}
\rho_{s,0} r_0^2\left(\frac{r}{r_0}\right)^{2-\gamma}\nonumber\\
&&\times\left[1+
\epsilon_z\frac{({\bf r}\cdot\hat{\bf n}_z)^2}{r^2}+
\epsilon_y\frac{({\bf r}\cdot\hat{\bf n}_y)^2}{r^2}\right].
\label{eq:phi_r}
\ea
The second term in the this equation represents the potential 
from a triaxial cluster.
From Poisson's equation we get the
following density distribution  (see e.g, \citealt{chandra69,2013degn.book.....M}):
\ba
\rho_s({\bf r})=\rho_{s,0}\left(\frac{r}{r_0}\right)^{-\gamma}
\left[1+\tilde{\epsilon}-
\tilde{\epsilon}_z\frac{({\bf r}\cdot \hat{\bf n}_z)^2\cdot}{r^2}
-\tilde{\epsilon}_y\frac{({\bf r}\cdot \hat{\bf n}_y)^2}{r^2}\right],\nonumber\\
\label{eq:rho_eps}
\ea
where
\ba
\tilde{\epsilon}_i=\epsilon_i\frac{\gamma(5-\gamma)}{(2-\gamma)(3-\gamma)},
\ea
with $i=y,z$ and
\ba
\tilde{\epsilon}=\frac{2}{(2-\gamma)(3-\gamma)} (\epsilon_y + \epsilon_z).
\ea

For $r\sim r_0$ the density distribution corresponds corresponds roughly to
an ellipsoidal mass distribution and it is useful to relate the quantities
above with axes ratios or ellipticities. 
In particular, for an axisymmetric potential ($|\epsilon_y|\ll |\epsilon_z|$)
and $\gamma=1$, approximations widely used in this paper,
we can express the axis ratios of the mass distribution as
\ba
\left(\frac{\tilde{a}}{\tilde{c}}\right)^2=
\left(\frac{\tilde{b}}{\tilde{c}}\right)^2=\frac{1+\tilde{\epsilon}-\tilde{\epsilon}_z}{1+\tilde{\epsilon}}=
\frac{1-\epsilon_z}{1+\epsilon_z},
\ea
where $\tilde{a}$, $\tilde{b}$, and $\tilde{c}$ correspond to the axis length
scales of the ellipsoidal mass distribution 
along $\hat{\bf n}_{x}$, $\hat{\bf n}_{y}$, and
$\hat{\bf n}_{z}$ respectively.
Thus, for $\epsilon_z>0$ the density distribution
represents an oblate spheroid ($\tilde{a}=\tilde{b}<\tilde{c}$), while 
$\epsilon_z<0$ it
represents a prolate spheroid ($\tilde{a}=\tilde{b}>\tilde{c}$).
Moreover, we can define the ellipticity\footnote{We define the ellipticity
as $1-q$ with $q$ being the ratio between the minor axis to the major axis
of the ellipsoid.  
Also, the eccentricity of the ellipsoid can be computed as
$\left[2|\epsilon_z|/(1+|\epsilon_z|)\right]^{1/2}$.} of the density distribution
for either oblate or prolate  distributions as
\ba
\mathcal{E}=1-\left(\frac{1-|\epsilon_z|}{1+|\epsilon_z|}\right)^{1/2}.
\label{eq:ellip}
\ea
Thus, $\epsilon_z\simeq0.05-0.5$ is equivalent to a distribution
with $\mathcal{E}\simeq0.05-0.4$, while the limiting 
case of a fully flattened distribution with $\mathcal{E}\to1$
corresponds to $|\epsilon_z|\to1$.
Throughout this paper, we will take the  
coefficients $\epsilon_y$ and $\epsilon_z$ 
as free parameters
and without loss of generality we will assume that
they are positive\footnote{The sign of $\epsilon_z$ 
only changes the direction of the nodal precession 
and has no effect for the statistical distributions 
of orbital elements of the binaries.}. 
We remark that formally the mass distribution 
in Equation (\ref{eq:rho_eps}) does not correspond to 
ellipsoidal mass distribution (unless $\gamma=0$) and 
the ellipticity is only defined locally for $r\sim r_0$ as
a reference.

Within the sphere of influence of the MBH the orbits are nearly Keplerian
and we can treat the force of the stellar cluster as a small 
perturbation that gradually torques the orbit, while not changing its energy--the 
so-called secular approximation.
As depicted in Figure \ref{fig:cluster},
we use the following coordinate system:
\ba
{\bf r}_{\rm in}&=&{\bf r}_1-{\bf r}_2\\
{\bf r}_{\rm out}&=&\frac{m_1{\bf r}_1+m_2{\bf r}_2}{m_1+m_2}-{\bf r}_{\rm BH},
\ea
with the MBH at the origin (${\bf r}_{\rm BH}=0$).
The Keplerian orbits are fully specified by the specific angular momentum
vectors ${\bf j}_i=\sqrt{1-e_i^2}\hat{\bf j}_i$ and the eccentricity vectors
${\bf e}_i=e_i\hat{\bf e}_i$ with $i=\{{\rm in},{\rm out}\}$, while
$\hat{\bf q}_i=\hat{\bf j}_i\times\hat{\bf e}_i$ completes the orthogonal
triad.
The equations of motion are fully described in the Appendix.


\section{Relevant timescales in the problem}
\label{sec:timescales}

The environment of a NC is a complicated, multi-timescale, system
and there are many ingredients coming into play when considering
the evolution of a binary.
In what follows, we quantify the most relevant (typically the shortest) 
timescales in the problem and compare this to the LK
timescale \citep[e.g.,][]{holman97}:
\begin{eqnarray}
\tau_{ \mbox{\tiny LK}}&=&  \frac{(m_1+m_2)}{M_{\rm BH}}
\frac{a_{\rm out}^3(1-e_{\rm out}^2)^{3/2}}{a_{\rm in}^3}\frac{2P_{\rm in}}{3\pi},\nonumber\\
&\simeq& 6\times10^4 \left(\frac{m_1+m_2}{20M\odot}\right)^{1/2}
\left(\frac{4\times10^6M_\odot}{M_{\rm BH}}\right)\nonumber\\
&&\times\left(\frac{a_{\rm out}[1-e_{\rm out}^2]^{1/2}}{0.1\mbox{ pc}}\right)^3
\left(\frac{a_{\rm in}}{10 \mbox{AU}}\right)^{-3/2} \mbox{ yr },
\label{eq:tau_lk}
\end{eqnarray}
where $P_{\rm in}$ is the period of the inner binary orbit.

We start in this section by discussing  some 
of the main dynamical processes  that can affect the evolution
of the inner orbit of the binaries as well as their orbit around the MBH.

For our galactic center we shall assume that  $M_{\rm BH}=4\times10^6M_\odot$
and that the spherical component of density profile (setting $\tilde\epsilon=\tilde\epsilon_z
=\tilde\epsilon_y=0$ in Eq. [\ref{eq:rho_eps}])
of the nuclear star cluster is given by
\ba
\rho_s(r)=0.8\times 10^5{M_\odot\rm pc^{-3}}\left(r\over 1{\rm pc}\right)^{-1.3}\ ,
\label{eq:rho_gc}
\ea
consistent with recent observational constraints 
\citep{2016ApJ...821...44F,2017arXiv170103817S}.

\subsection{Evolution of the binary orbit around the MBH}

\subsubsection{Cluster-induced nodal and apsidal precession} \label{Nprec}

A key new ingredient in our paper is the 
effect that the cluster gravitational potential 
has on the binary orbit around the MBH.
There are two main dynamical effects that we describe next.

\paragraph{Nodal precession and possible changes
of $j_z$}
Provided that the cluster is not fully spherically symmetric 
($\epsilon_z>0$ in Equation \ref{eq:phi_r}), then ${\bf j}_{\rm out}$
will precess around ${\bf \hat{n}}_z$. 
From Equations (\ref{eq:jout_prec}) and (\ref{eq:tau_sc})
we find that the characteristic
timescale for this nodal precession is
\ba 
\tau_{\rm sc,z}&\equiv&\frac{\tau_{\rm sc}}{\epsilon_z}=
\frac{1 }{\epsilon_z G P_{\rm out} \rho_s(a_{\rm out})}\nonumber\\
&\simeq&1.4\times 10^{6}\mbox{ yr }\left(\frac{\epsilon_z}{0.1}\right)^{-1}
\left(\frac{\rho_s(a_{\rm out})}{10^6~M_\odot\mbox{pc}^{-3}}\right)^{-1}\nonumber\\
&&\left(\frac{M_{\rm BH}}{4\times10^6M_\odot}\right)^{1/2}
\times\left(\frac{a_{\rm out}}{0.1\mbox{pc}}\right)^{-3/2},
\label{eq:nodal}
\ea
and for a generic density profile we 
have that $\tau_{\rm sc,z}\propto (a_{\rm out})^{\gamma-3/2}$.
Since $\tau_{ \mbox{\tiny LK}}\propto a_{\rm out}^3$, we generally 
have that $\tau_{ \mbox{\tiny LK}}\ll\tau_{\rm sc,z}$ for small
$a_{\rm out}$, while the opposite regime is reached for large 
$a_{\rm out}$. Thus, there is region in which  
$\tau_{ \mbox{\tiny LK}}\sim\tau_{\rm sc,z}$.

Assuming the density distribution
in Equation (\ref{eq:rho_gc})
we can express the nodal precession rate
($=1/\tau_{\rm sc,z}$) in units of the
LK timescale as
\ba
\epsilon_{\rm sc,z}&&\equiv\epsilon_z\frac{\tau_{\mbox{\tiny LK}}}{\tau_{\rm sc}}
\simeq0.07~\left(\frac{\epsilon_z}{0.1}\right)
\left(\frac{10\mbox{ AU}}{a_{\rm in}}\right)^{3/2}\nonumber \\
&&\times\left(\frac{m_1+m_2}{20M_\odot}\right)^{1/2}
\left(\frac{a_{\rm out}}{0.1\mbox{ pc}}\right)^{3.2}(1-e_{\rm out}^2)^{3/2}.
\label{eq:omega_sz}
\ea
As we will show in \S\ref{sec:dynamics}, this ratio of timescales
dramatically changes the behavior of the inner binary and
for $\epsilon_{\rm sc,z}\gtrsim0.1$ significant changes of 
$j_z=\sqrt{1-e_{\rm in}^2}\cos i_{\rm in, out}=\sqrt{1-e_{\rm in}^2}~
\hat{\bf j}_{\rm in}\cdot\hat{\bf j}_{\rm out}$
occur, leading to new interesting phenomena.
Geometrically the changes in  $j_z$ are due to changes in
$ i_{\rm in, out}$ from the fast precession of 
${\bf j}_{\rm out}$ around ${\bf \hat{n}}_z$. 

Finally, we note that in triaxial clusters ($\epsilon_y>0$ in Equation \ref{eq:phi_r}) the
orbit will also precess around ${\bf \hat{n}}_y$ with characteristic 
timescale $\tau_{\rm sc,y}\equiv \tau_{\rm sc}/\epsilon_y$.

\paragraph{Apsidal precession and suppression 
of the octupole-level dynamics}
Another important effect from the cluster is the orbit apsidal 
precession induced by the dominant 
spherical mass distribution. 
This important effect was neglected in most previous work 
\citep{2015ApJ...799..118P,2016MNRAS.460.3494S}.
For the purposes of our model, the main dynamical effect
is to quench the non-axisymmetric  contribution (octupole-level or higher 
order terms) from the outer orbit. In other words, if the outer obit precesses
fast compared to $\tau_{\mbox{\tiny LK}}$, its time-averaged potential will be 
effectively axisymmetric, even for $e_{\rm out}>0$.

To quantify this effect, we define the ratio between the
octupole timescale, $\sim \tau_{\mbox{\tiny LK}}/\epsilon_{\rm oct}$ with
$\epsilon_{\rm oct}$ from Equation (\ref{eq:oct}), and
the precession due to the spherical cluster using 
from Equation  (\ref{eq:tau_sc}) as 
\ba 
\epsilon_{\rm oct,sc}&\equiv& \epsilon_{\rm oct}^{-1}
\frac{\tau_{\mbox{\tiny LK}}}{\tau_{\rm sc}}\simeq1.2\left(\frac{10\mbox{ AU}}{a_{\rm in}}\right)^{5/2}
\left(\frac{a_{\rm out}}{0.02\mbox{ pc}}\right)^{4.2}\nonumber\\
&\times&
\left(\frac{m_1+m_2}{20M_\odot}\right)^{1/2}
\left(\frac{m_1+m_2}{m_1-m_2}\right)
\frac{(1-e_{\rm out}^2)^{5/2}}{e_{\rm out}},
\label{eq:oct_sc}
\ea
where we assumed a cluster density
profile given by Equation (\ref{eq:rho_gc}).
For $\epsilon_{\rm oct,sc}\gg1$ 
we expect that the octupole-level perturbations 
do not contribute significantly 
to the excitation of large eccentricities of the binary.

We note that regardless of the cluster-driven apsidal precession,
the octupole-level perturbations are expected 
to be only relevant if $\epsilon_{\rm oct}\gtrsim10^{-3}$. This condition
can be written as
$a_{\rm in}/[10\mbox{AU}]\gtrsim a_{\rm out}/[0.02\mbox{pc}]$
for $m_1\gg m_2$ and $e_{\rm out}\sim0.5$.
In order to compare which condition dominates at suppressing the
octupole,  we write the condition $\epsilon_{\rm oct,sc}\gtrsim1$ as
$a_{\rm in}/[10\mbox{AU}]\gtrsim  (a_{\rm out}/[0.02\mbox{pc}])^{8.4/5}$.
Thus, given that the latter condition scales more steeply with $a\out$
we expect that the cluster is the dominant source of suppression
for $a\out\gg0.02$ pc, while the condition from triple dynamics
dominates for $a\out\ll0.02$ pc.
In either case, we expect that the octupole is highly suppressed
for $a\out\gg0.02$ pc.

We conclude that the octupole-level potential from the outer orbit is expected
to have a negligible contribution at exciting extreme eccentricities 
at separations of $a_{\rm out}\sim0.1-1$ pc.
It can, however, be an important ingredient in the dynamics 
inside $a_{\rm out}\lesssim0.01$ pc.

\subsubsection{Two-body and resonant relaxation}
In the dense stellar environment of a NC, stellar orbits 
can be significantly affected by their frequent interactions with field stars,
which cause their orbital energy and angular momentum to random walk.
The timescale associated with this process, referred to as two-body relaxation,
is \citep{1987degc.book.....S}:
\begin{eqnarray}
{\tau_{\rm 2b} }&\approx&~ 10 \tau_{\rm evap}\times
\left(\ln \Lambda\over \ln \Lambda' \right)
{a_{\rm in}\sigma^2\over G \left(m_1+m_2\right)^{-1}}\nonumber\\
&\approx&  ~10
\tau_{\rm evap}
\left({10\over 2} {\ln \Lambda\over \ln \Lambda'} \right)
\left(m_1+m_2\over 20M_\odot\right)^{-1}\nonumber \\
&& \times
\left(\sigma\over 100\rm km\ s^{-1}\right)^{2} 
\left(a_{\rm in}\over 10\rm AU\right) ,
\end{eqnarray}
where $\Lambda'\sim M_{\rm BH}/m$
and $\Lambda\sim (m_1+m_2)/m$ are
the Coulomb logarithms with $m$ the typical mass
of the stars in the cluster. 
Note that $\Lambda'\gg\Lambda$, 
since encounters on all scales contribute to the relaxation process.
This timescale is clearly much longer than 
the other relevant timescales in the problem (e.g., $\tau_{\rm coll};\ \tau_{\rm LK}$). 
A massive binary will also segregate to the center via
dynamical friction. The associated timescale 
is $\approx \tau_{\rm 2b}m/(m_1+m_2)$; an expression roughly 
valid for a steep density profile cusp \citep{2016arXiv161106573D}. Thus,  
 dynamical friction acts on a timescale which is typically much longer than
$\tau_{\rm coll}$, while it can become of order $\tau_{\rm evap}$ for
relatively massive binaries ($\gtrsim 10\ M_\odot$), with 
small orbital semi-major axis $a_{\rm in}\lesssim 1\rm AU$, and 
that reside at  large galactocentric distances ($\sigma \lesssim 100\rm km\ s^{-1}$).
Such binaries are, however, unlikely to be affected by LK-like oscillations
given the strong quenching  
caused by relativistic precession (see \S\ref{rpre}).

Based on our discussion, we conclude that  collisional   
(gravitational-encounter-driven) relaxation processes are likely 
to become important only after the binaries have already evaporated from the NC. However, very near a MBH orbital angular momenta can evolve as a consequence of 
 resonant relaxation \citep{1996NewA....1..149R,2006ApJ...645.1152H}.
Due to the finite number of stars in the vicinity of a MBH and because 
their Keplerian orbits 
maintain their orientation fixed in space for a long time, 
torques acting on a test star from all other stars do not cancel exactly.
The resulting $\sqrt{N}$ deviation from  spherical symmetry induces a net 
residual torque on the test star orbit.

The frequency of the orbital plane precession
associated with resonant relaxation is:
$\tau_\mathrm{RR} \approx \tau_\mathrm{sc}\sqrt{N}$
with $N$ approximately the number of stars within a sphere of radius $a_\mathrm{out}$.
Comparison with the timescale associated with the nodal precession
in a triaxial potential (Eq. [\ref{eq:nodal}]) shows that 
$\tau_\mathrm{RR}/\tau_\mathrm{sc,z}\approx \epsilon_z \sqrt{N}$.
From Equation (\ref{eq:rho_gc}) setting the typical stellar mass in the nuclear cluster 
equal to $0.5M_\odot$ and $\gamma=1$,
we have $N\approx 10^6 \left(a_\mathrm{out}/\mathrm{1pc} \right)^{2}$. It follows that
precession due to the aspherical cluster
will dominate over resonant relaxation at radii $a_\mathrm{out}\gtrsim 10^{-3}\mathrm{pc}/\epsilon_z$.
If we set $\epsilon_z=0.1$, then resonant relaxation can be  ignored at $a\out\gtrsim0.01\mathrm{pc}$.

The evolution of binaries due to resonant relaxation couple with the LK process
is a complex phenomenon and will be investigated in future work.
Here we limit to notice that there is clearly a connection between the evolution
of orbits in non-spherical clusters and due to resonant-relaxation. 
As also noted in \citet{merritt2011} and \citet{MV11},
during the coherent phase of resonant relaxation, the gravitational potential
from the star cluster can be represented in terms
of a multipole expansion; if the lowest-order non-spherical 
terms in that expansion coincide with the
potential generated by a  triaxial cluster,
then the behavior of orbits in the coherent regime would
be the same as that  derived below for orbits in a triaxial nucleus.
We  therefore expect
the reorientation of orbital planes due to resonant relaxation to result 
into an evolution similar to the one described  below in the case of 
a non-spherical star cluster.

\begin{figure*}
\center
\includegraphics[height=7.5cm]{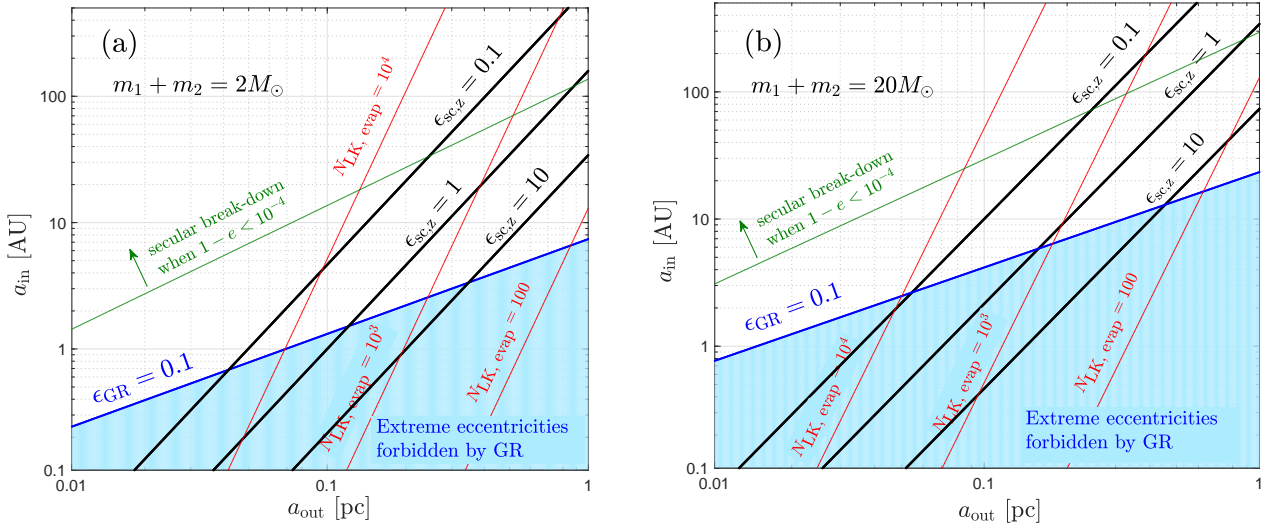}
\caption{Regions in $a_{\rm in}-a_{\rm out}$ space where
the mechanism described in this paper operates most
efficiently at exciting large eccentricities, 
which corresponds to $\epsilon_{\rm sc,z}\sim0.1-10$ (black lines) in
Equation  (\ref{eq:omega_sz})  (see Figure 
\ref{fig:population_epsz} for reference).
The shaded region  corresponds to $\epsilon_{\rm GR}<0.1$ (Equation \ref{eq:eps_GR}), where relativistic precession in the inner 
binary quenches the extreme
excitation of eccentricities.
The red lines indicate the number of LK cycles allowed before 
the binary evaporates $N_{ \mbox{\tiny LK, evap}}$ 
(Eq. [\ref{eq:Nevap}])
and the green line corresponds to the limit above which 
the secular approximation breaks down 
when $1-e\lesssim10^{-4}$ from Equation (18) in
\citet{AMM2014}.
The cluster potential in Equation (\ref{eq:phi_r}) is assumed 
to be axisymmetric 
with $\epsilon_{\rm z}=0.1$ and the spherical component of the density 
profile is given by the Equation (\ref{eq:rho_gc}).
{\it Panel a}: total mass of the binary of $2M_\odot$.
{\it Panel b}: total mass of the binary of $20M_\odot$.}
\label{fig:timescales}
\end{figure*}

\subsection{Evolution of the binary inner orbit}

\subsubsection{Binary evaporation} 
\label{sec:evap}

Due to the large velocity dispersion near a MBH,
most binaries in a nuclear cluster are very soft, 
meaning that $|E_b|/(m\sigma^2)\ll1$, where $m$ is the 
typical stellar mass, $E_b=-G(m_1+m_2)^2/(2a_{\rm in})$,
and $\sigma$ is the 1D stellar velocity dispersion.
Encounters with field stars will make binaries softer
and disrupt them \citep{H75}.

The timescale on  which a binary experiences an encounter 
with a field star  is $t_{\rm coll}=(\nu\Sigma\sqrt{3}\sigma)^{-1}$, 
where  $\nu$ is the  number density of stars,
and $\Sigma=\pi r_{\rm coll}^2 \left[1+2G(m_1+m_2+m)/(\sqrt{3}\sigma)^2r_{\rm coll} \right]$
is the interaction cross-section for
periapsis distances $\leq r_{\rm coll}$.
The timescale for a binary to experience a close 
encounter with a field star is then (e.g., \citealt{BT08}):
\begin{eqnarray}
\tau_{\rm coll}&=&{1\over ({\pi /4})\nu \sigma a_{\rm in}^2(1+\Theta)}\approx
 {5\times 10^6\over(1+\Theta) }  \left(\nu \over10^6 \rm pc^{-3} \right)^{-1} \\
&&
\times \left(\sigma\over
100\rm km\ s^{-1} \right)^{-1} \left(a_{\rm in} \over 10{\rm AU}\right)^{-2} 
\rm yr
\label{eq:tau_coll}
\end{eqnarray}
with
\begin{eqnarray}
\Theta={{4 } {\rm G} (m_1+m_2+m)\over (\sqrt{3}\sigma)^2 a_{\rm in}}=0.08
\left( {m_1+m_2+m\over 20M_\odot}\right)
\\
 \left(\sigma\over
100\rm km\ s^{-1} \right)^{-2} \left(a_{\rm in}\over 10 \rm AU \right)^{-1}. \nonumber
\end{eqnarray}
In the expressions above, we set $r_{\rm coll}=a_{\rm in}/2$. 
In more distant encounters the effect
of the field particle on one of the binary components almost cancels 
the effect on the other component \citep[e.g.,][]{H75}.
While more distant encounters can still (secularly) 
change the eccentricity and orientation of the binary, 
we reserve a careful treatment of 
this potentially important effect to future work.

Most  relevant  for  our  work  is   the  
ratio between the collision timescale and the Lidov-Kozai
timescale:
\begin{eqnarray}
N_{ \mbox{\tiny LK, coll}}&&\equiv{\tau_{\rm coll}\over \tau_{ \mbox{\tiny LK}}}\approx 
{80\over(1+\Theta) }  \left(1-e_{\rm out}^2 \right)^{-3/2} \nonumber
\\
\times&& \left(\frac{a_{\rm in}}{10~\mbox{AU}}\right)^{-1/2}\left(\frac{a_{\rm out}}{0.1~\mbox{pc}}\right)^{-3} 
\left(\frac{m_1+m_2}{20 M\odot}\right)^{-1} 
 \nonumber \\
\times&& \left( M_{\rm BH}\over 4\times10^6M_\odot \right) \left(\nu \over10^6 \rm pc^{-3} \right)^{-1}   \left(\sigma\over 100\rm km\ s^{-1} \right)^{-1},\nonumber\\
\label{eq:Ncoll}
\end{eqnarray}
as long as $N_{ \mbox{\tiny LK}}\gtrsim 1$ the effect of encounters 
only becomes important after many LK cycles.

Strictly speaking, since we neglect the effect of encounters with field stars in our
calculations, on timescales that are longer than $\tau_{\rm coll}$ 
our treatment might  not longer be valid.
However, the timescale over which 
 impulsive encounters will fully disrupt a
 binary is longer than $\tau_{\rm coll}$. 
After only multiple encounters ($N_{ \mbox{\tiny LK}}>1$),  the orbital elements
of the binary might change significantly 
and eventually lead
to the evaporation of the binary. The evaporation time
can be estimated as \citep{BT08}:
\begin{eqnarray}
\tau_{\rm evap}&=&{(m_1+m_2)\sigma\over 16\sqrt{\pi} G a_{\rm in} m^2 \nu  
\ln \Lambda}
= 3\times 10^8 \left(\ln \Lambda \over 2\right)^{-1}\nonumber\\
&&\times\left(m_1+m_2\over 20M_\odot\right) 
 \left( M_{\rm BH}\over 4\times 10^6 M_\odot\right)^{-1}
\left(\sigma\over 100\rm km\ s^{-1}\right)  \nonumber
\\
&&\times \left(a_{\rm in} \over10\rm AU \right)^{-1} 
\left(m\nu \over 5\times 10^5 M_\odot\ {\rm pc^{-3}}\right)^{-1}
\rm yr.
\label{eq:tau_evap}
\end{eqnarray}
The  ratio between the evaporation timescale and the Lidov-Kozai
timescale for the galactic center 
($m\nu=\rho_s$ from Eq. [\ref{eq:rho_gc}])
and assuming $\sigma=280\rm km\ s^{-1}$$\sqrt{0.1\mbox{pc}/a\out}$ 
(e.g., \citealt{KT11}) is
\begin{eqnarray}
N_{ \mbox{\tiny LK, evap}}&\equiv& {\tau_{\rm evap}\over\tau_{\rm LK}}
\approx 4400 \left(a_{\rm in}\over 10\rm AU\right)^{1/2}
\left(m_1+m_2\over 20M_\odot\right)^{1/2} \nonumber \\
&&\times \left(a_{\rm out}\over \rm 0.1pc \right)^{-2.2}\ ,
\label{eq:Nevap}
\end{eqnarray} 
which is an estimate of the number of LK cycles that the
binary can undergo before it fully 
``evaporates'' due to encounters with stars in the NC.

We remark that the effect from encounters
in the binary orbit is a
stochastic process and  the statistical fluctuations governing it
 are not taken into account in our work
 (see \citealt{BML17} for a recent work that incorporates
 these stochastic processes). 
Instead we study how the results behave as a function of
the number of cycles (see \S\ref{sec:Nmax}).

\subsubsection{Relativistic perturbations} \label{rpre}
Two compact-objects reaching sufficiently close to each other lose 
energy by emitting gravitational wave (GW) radiation.

The merger timescale mediated by gravitational radiation
is \citep{1964PhRv..136.1224P}:
\ba
\tau_{\rm GW}&=&\frac{3}{85}\frac{a}{c}\left(\frac{a_{\rm in}^3c^6}{G^3m_1m_2(m_1+m_2)}\right)\left(1-e_{\rm in}^2\right)^{7/2},\\
\ea
which at very large eccentricities ($1-e^2\simeq2[1-e]$), relevant for our 
work, can be written as
\ba
\tau_{\rm GW}&\simeq&4.1\times10^4\mbox{  }\left(\frac{m_1m_2(m_1+m_2)}{2\times10^3M_\odot^3}\right)^{-1}
\nonumber\\
&&\times\left(\frac{a_{\rm in}}{10\mbox{~AU}}\right)^{1/2}
\left[\frac{a_{\rm in}(1-e_{\rm in})}{0.001\mbox{~AU}}\right]^{7/2}
\rm yr.\nonumber\\
\ea
In order to have a rapid merger such that the binary orbit
shrinks significantly in one LK cycle we require 
the change in semi-major axis due to GW loses, 
$\tau_{\rm GW}$, to be shorter than
the change in the periapsis distance $a_{\rm in}(1-e_{\rm in})$ due to
MBH, which at higher eccentricities is
$\simeq\sqrt{1-e_{\rm in}^2}\tau_{ \mbox{\tiny LK}}$.
For BH binaries with $m_1+m_2=10M_\odot$ initially at $a_{\rm in}=10$ AU, 
this condition for rapid merger reads:
\ba
1-e_{\rm merger}\lesssim3\times10^{-5}\times
\left(\frac{a_{\rm out}(1-e_{\rm out}^2)^{1/2}}{0.1\mbox{ pc}}\right)\ .
\label{eq:e_merge}
\ea

It is well-known that LK oscillations can be quenched 
by additional sources of 
apsidal precession of the inner binary orbit \citep[e.g.,][]{blaes02}. 
This can prevent the binary from reaching the high eccentricities
 required for a rapid merger to occur. 
We parametrize the importance 
of relativistic precession by the ratio between the LK timescale and the GR 
precession timescale as (e.g., \citealt{FT07})
\ba
\epsilon_{\rm GR}\equiv\frac{ \tau_{ \mbox{\tiny LK}}}{\tau_{\rm GR}}&=&
\frac{4G(m_1+m_2)^2a_{\rm out}^3(1-e_{\rm out}^2)^{3/2}}
{c^2a_{\rm in}^4M_{\rm BH}}\nonumber\\
&\simeq& 3\times 10^{-3}\left(\frac{m_1+m_2}{20M_\odot}\right)^2
\left(\frac{a_{\rm in}}{10\mbox{~AU}}\right)^{-4}\nonumber\\
&&\left(\frac{a_{\rm out}\sqrt{1-e_{\rm out}^2}}{0.1\mbox{~pc}}\right)^3.
\label{eq:eps_GR}
\ea
As we show below in Section \S\ref{sec:eps_GR}, the eccentricities do no reach 
$1-e\lesssim10^{-3}$ ($1-e\lesssim10^{-5}$) for $\epsilon_{\rm GR}>0.1$
($>0.01$).

\subsection{Available phase-space for
eccentricity excitation.}

In Figure \ref{fig:timescales} we show the regions in $a_{\rm in}-a_{\rm out}$
space  where our model can potentially excite large eccentricities for
two values of $m_1+m_2$.

Provided that the binary can undergo several LK cycles before evaporating
($N_{ \mbox{\tiny LK,evap}}\gg1$; Eq. [\ref{eq:Nevap}]), the main limiting timescale to achieve 
extreme eccentricities is the relativistic precession.
 For reference, we show the shaded region that 
 corresponds to $\epsilon_{\rm GR}<0.1$ 
(Equation [\ref{eq:eps_GR}]). 
As we show later in \S\ref{sec:eps_GR}
we require that $\epsilon_{\rm GR}\lesssim0.1$ 
($\lesssim0.01$) to reach 
$1-e_{\rm max}\lesssim10^{-3}$ ($1-e_{\rm max}\lesssim10^{-5}$).

The regions where the effect from the cluster
can potentially enhance the LK mechanism
are indicated in diagonal black lines.
As we will show, the cluster mainly enhances the excitation of
large eccentricities when the timescale at which it forces
the nodal precession of the 
outer orbit becomes comparable
to the LK timescale.
We parametrize this 
with the ratio of timescales 
$\epsilon_{\rm sc,z}\equiv\epsilon_z\tau_{\mbox{\tiny LK}}/\tau_{\rm sc}$ 
defined in Equation  (\ref{eq:omega_sz}). 
The black diagonal lines indicate $\epsilon_{\rm sc,z}=\{0.1,1,10\}$
and should serve only as reference since they  assume a cluster with a 
particular density distribution and shape ($\epsilon_z=0.1$ or ellipticity
of $\sim0.1$).

From this Figure we conclude that our mechanism
is most likely to play a significant role for binaries at separations of
$a_{\rm out}\gtrsim0.05$ pc from the MBH and with 
semi-major axes $a_{\rm in}\gtrsim1$ AU.
However, we note that as $a_{\rm out}$ increases,
the number of LK cycles allowed before the binary evaporates
due to encounters from passing stars 
$N_{ \mbox{\tiny LK, evap}}$ 
in (Eq. [\ref{eq:Nevap}], red lines) decreases.
Since the behavior at $\epsilon_{\rm sc,z}\sim1$ is often chaotic, the
number of cycles limits the phase-space exploration
to reach $e\to1$ and, therefore, the merger rates. 

Finally we show the region where our secular equations of 
motion in Appendix are expected to become invalid 
when the orbit loses enough angular momentum
($1-e\lesssim 10^{-4}$) due to the break down
of the secular approximation (region above green line from
Equation 18 in \citealt{AMM2014}).
When this occurs the orbital elements of the inner binary can
change in dynamical timescales due to the torque from the outer
orbit, while the semi-major axes remains fixed. 
This behavior is not captured by our secular code, but we expect
that it does not invalidate our conclusion that the merger rates
are high when $\epsilon_z\sim1$ because the angular momentum
changes in dynamical timescales generally lead to extremely
large values of the eccentricities, as previously noted
by several authors
(e.g., \citealt{2012ApJ...757...27A,KD12,ASTA2014,AMM2014}).

\begin{figure*}
\center
\includegraphics[width=\textwidth]{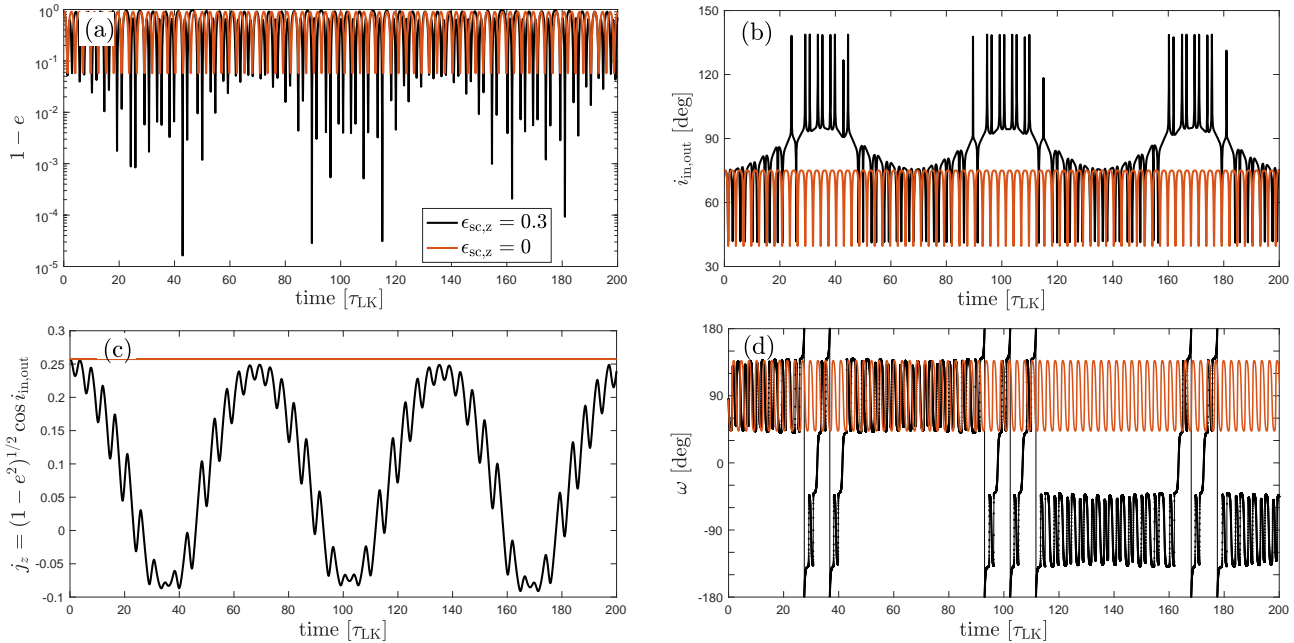}
\caption{Evolution of a binary in a cluster with $\epsilon_{\rm sc,z}=0.3$ in Equation 
(\ref{eq:omega_sz}) (black lines), compared to the case of a non-existing 
(or fully spherical) cluster, $\epsilon_{\rm sc,z}=0$ (red lines).
The initial conditions relative to the outer (binary-MBH system) 
are $i_{\rm in,out}=75^\circ$, $\omega=90^\circ$, $\Omega=0$, $e=0.01$, while outer
orbit is initially oriented relative to the cluster axis with 
$i_{\rm out,z}=85^\circ$ and  $\Omega_{\rm out,z}=180^\circ$,
and have $e_{\rm out}=0$.
{\it Panel (a):} eccentricities.
{\it Panel (b):} inclination between inner and outer orbit $i_{\rm in,out}$ (angle between
$\hat{\bf j}_{\rm in}$ and $\hat{\bf j}_{\rm out}$).
{\it Panel (c):} $j_{\rm z}=(1-e^2)^{1/2}\cos i_{\rm in,out}$, which due to the nodal
precession of ${\bf j}_{\rm out}$ is no 
longer conserved (see Equation [\ref{eq:jz_evol}]).
{\it Panel (d):} argument of periapsis $\omega$.}	
\label{fig:example1}
\end{figure*}

\section{Dynamical behavior}
\label{sec:dynamics}

We describe the dynamical behavior that arises from including the
effect from the cluster potential in the dynamics of binaries.
We compare our results with those from evolving the binaries
in the absence of a cluster (or a having a fully spherical cluster),
in which case the dynamics reduces to the well-known 
Lidov-Kozai mechanism at quadrupole-order
(expansion up to $a_{\rm in}^2/a_{\rm out}^3$).

We focus on the regime in which the main effect of the
cluster is to force the outer orbit to undergo nodal 
precession (see Figure \ref{fig:cluster}
for reference), but not change the magnitude of
its angular momentum substantially.
We remark that the model in the Appendix (Eqs. [\ref{eq:edot}]-[\ref{eq:jout_evol}])
is not subject to this simplification and it can handle strong variations of
the angular momentum of the outer orbit due to the triaxial potential.

\subsection{Simplified two-degree of freedom model} 
\label{sec:simple}

We consider the regime in which the outer orbit has enough 
angular momentum
and/or the cluster is not sufficiently flattened (or triaxial), so that the magnitude 
of the outer
orbit angular momentum $j_{\rm out}=(1-e_{\rm out}^2)^{1/2}$ does not change substantially. 
In practice, for mathematical simplicity, we consider the following limiting 
regime:
\begin{itemize}
\item  nearly circular outer orbit: $e_{\rm out}\ll1$;
\item weakly axisymmetric cluster potential: $\epsilon_y\ll\epsilon_z\ll1$ in Equation
(\ref{eq:phi_r}).
\end{itemize}

In this limit, the  Equation (\ref{eq:eout_evol}) becomes
\ba
 \frac{d {\bf e}_{\rm out}}{dt}&=& 
 \tau_{\rm sc}^{-1}\left(
 {\bf e}_{\rm out}\times{\bf j}_{\rm out}\right),
\ea
where $\tau_{\rm sc}$ is the timescale of variations of the 
outer orbit due to the spherical component of the stellar cluster given in
Equation (\ref{eq:tau_sc}).
This last equation implies that ${\bf e}_{\rm out}$ only undergoes apsidal
precession and $e_{\rm out}$ remains small.
In turn, this small  $e_{\rm out}$ 
implies that Equation (\ref{eq:jout_evol}) becomes:
\ba
\frac{d \hat{\bf j}_{\rm out}}{dt}&=&\tau_{\rm sc}^{-1}\sum_{i=y,z} \epsilon_i
\left(\hat{\bf n}_i\cdot \hat{\bf j}_{\rm out}\right)\hat{\bf j}_{\rm out}\times
 \hat{\bf n}_i
 \label{eq:jout_prec}
\ea
meaning that $\hat{\bf j}_{\rm out}$ is simply
undergoing nodal precession around  $\hat{\bf n}_y$
and $\hat{\bf n}_z$ with rates proportional to $\epsilon_y$
and $\epsilon_z$, respectively.
Simplifying things further, for an axisymmetric cluster ($\epsilon_y\ll\epsilon_z$)
we have 
\ba
\frac{d \hat{\bf j}_{\rm out}}{dt}&=&\tau_{\rm sc}^{-1}\epsilon_z
\left(\hat{\bf n}_z\cdot \hat{\bf j}_{\rm out}\right)\hat{\bf j}_{\rm out}\times
 \hat{\bf n}_z
\ea
meaning that $\hat{\bf j}_{\rm out}$ simply precesses
around $ \hat{\bf n}_z$ at a constant rate 
$-\epsilon_z/\tau_{\rm sc}\left(\hat{\bf n}_z\cdot \hat{\bf j}_{\rm out}\right)$.
Conveniently, we can move to the  rotating
coordinate system that co-precesses with 
the outer orbit
and write the potential in Equation (\ref{eq:phi_full})
due to the MBH and the
star cluster with normal vector  $\hat{\bf n}_z$ as
(e.g., \citealt{TY14})
\ba
\frac{\bar{\Phi}_{\rm rot}}{\phi_0}&=&
\frac{1}{6}-e^2 -\frac{1}{2}({\bf {j}}\cdot \hat{\bf j}_{\rm out})^2+
\frac{5}{2}  ({\bf e}\cdot \hat{\bf j}_{\rm out})^2\nonumber\\
&+&\frac{(\epsilon_z/\tau_{\rm sc})\sqrt{G(m_1+m_2)a_{\rm in}}}{\phi_0}
\left(\hat{\bf n}_z\cdot \hat{\bf j}_{\rm out}\right)
\left({\bf {j}}\cdot\hat{\bf n}_z\right),\nonumber\\
\label{eq:phi}
\ea
where
\ba
\phi_0=\frac{3GM_{\rm BH} a_{\rm in}^2}{4a_{\rm out}^3(1-e_{\rm out}^2)^{3/2}}.
\ea
We note that the dimensionless term multiplying 
$\left(\hat{\bf n}_z\cdot \hat{\bf j}_{\rm out}\right)
\left({\bf {j}}\cdot\hat{\bf n}_z\right)$ is
simply $\epsilon_{\rm sc,z}\equiv\epsilon_z \tau_{\mbox{\tiny LK}}/\tau_{\rm sc}$
defined in Equation (\ref{eq:omega_sz}).
This parameter will be used throughout this paper and, as we will show, it critically 
determines the dynamical behavior of the inner binary.

In order to make the connection with a ``perturbed" Lidov-Kozai mechanism
more familiar to the reader, we write the potential in Equation 
(\ref{eq:phi}) in terms of orbital elements. By doing this,
we arrive at the following dimensionless two-degree of 
freedom Hamiltonian:
\ba
\tilde{\mathcal{H}}_{\rm rot}&=&-\frac{1}{3}-\frac{1}{2}e^2+
\left(\frac{1}{2}+2e^2-\frac{5}{2}e^2\cos^2\omega\right)
\sin^2i\nonumber\\
&+&\epsilon_{\rm sc,z}(1-e^2)^{1/2}\bigg[
\cos i\cos^2 i_{\rm out, z}'\nonumber\\
&+&\frac{1}{2}\sin i\sin 2i_{\rm out,z}' 
\cos\left(\Omega-\Omega_{\rm out,z}'\right)\bigg],
\label{eq:H_BHs}
\ea
where we assume that $m_1,m_2\ll M_{\rm BH}$, so we 
can define our coordinate system
in the rotating frame as
$\hat{\bf x}\equiv\hat{\bf e}_{\rm out}$ and 
$\hat{\bf z}\equiv\hat{\bf j}_{\rm out}$. 
For the sake of brevity, the variables of the inner orbit have no sub-index
and all the angles are defined relative to the outer orbit
(e.g., $i\equiv i_{\rm in,out}$).
The angles $i_{\rm out,z}'$ and $\Omega_{\rm out,z}'$ are the inclination and the 
longitude of the ascending node 
of the cluster symmetry
axis $\hat{\bf n}_z$ relative to the outer orbit in the rotating frame, both of 
which are constant in this frame.
Note that this Hamiltonian can be re-written in terms of 
the dimensionless\footnote{The actual Delaunay's variables have the momenta
$j$ and $j_z$ multiplied by $m_1m_2/(m_1+m_2)\sqrt{G(m_1+m_2)a}$.} 
canonical coordinate-momentum pairs:
$\left\{\omega,j=\sqrt{1-e^2}\right\}$ and 
$\left\{\Omega,j_z=\sqrt{1-e^2}\cos i\right\}$.

This Hamiltonian  shows explicitly that the momentum 
coordinate $j_z={\bf j}\cdot \hat{\bf j}_{\rm out}=\sqrt{1-e^2}\cos i$ 
can change due to the cluster potential
and it varies as:
\ba 
\frac{d j_z}{d\tau}=-\frac{d\tilde{\mathcal{H}}_{\rm rot}}{d\Omega}
&=&\frac{(1-e^2)^{1/2}}{2}\epsilon_{\rm sc,z}\sin i\nonumber\\
&\times&\sin 2i_{\rm out,z}' 
\sin\left(\Omega-\Omega_{\rm out,z}'\right),
\label{eq:jz_evol}
\ea
where the unit time is given in Lidov-Kozai timescales, $\tau=t/\tau_{\mbox{\tiny LK}}$.
This variation of $j_z$ is a crucial new ingredient in the dynamics
of the stellar binary and as we will show in the next sections it can lead
to the excitation of extremely high
eccentricities for a wide range of initial
conditions.

We note that this effect of forcing  $j_z$  is similar the effect from high-order 
terms in the triple interaction potential (octupole and higher-order, see \citealt{naoz16}), 
but unlike the three-body interaction, 
this forcing is not limited to happen in timescales much longer than 
$\tau_{\mbox{\tiny LK}}$\footnote{The 
octupole timescale is longer than $\tau_{\mbox{\tiny LK}}$ by a factor
$\sim a_{\rm out}(1-e_{\rm out}^2)/(a_{\rm in}e_{\rm out})$, while the timescale
from higher-order terms are even longer than the octupole
timescale.} 
(i.e., $\epsilon_{\rm sc,z}$  is not necessarily a small
parameter). In other words, for reasonable parameters of the
triple system and the cluster potential,  $\epsilon_{\rm sc,z}$ can 
take a wide range of values, including 
values much larger than unity.

\begin{figure*}\centering
\includegraphics[angle=0,width=18.5cm]{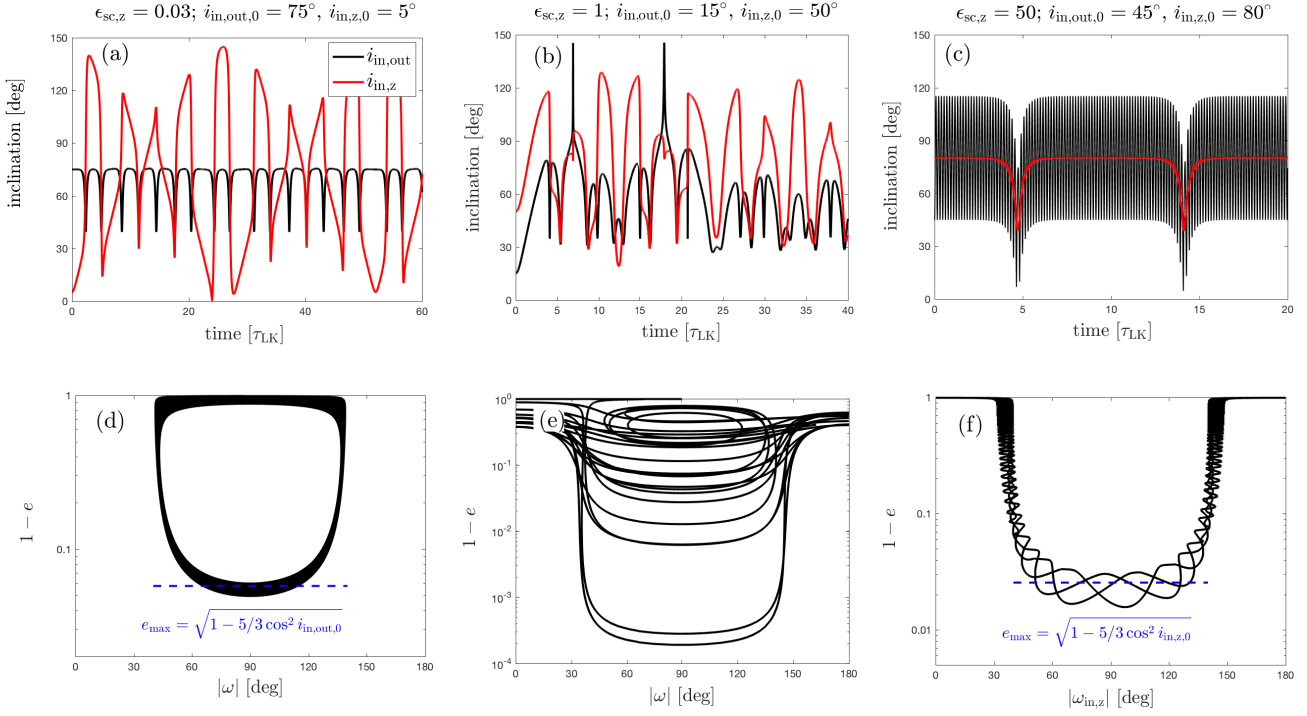}
\caption{Examples showing the evolution of the binaries eccentricities
and inclinations in three dynamical regimes of small to large $\epsilon_{\rm z}$ described in
Section \S\ref{sec:dynamics}. 
In these integrations we set $\omega=90^\circ$, $\Omega=0$, $e=0.01$, $e_{\rm out}=0.01$,
and $\epsilon_{\rm z}=0.1$, while varying $\epsilon_{\rm sc}$ to get
different values of  $\epsilon_{\rm sc,z}=\epsilon_{\rm z}\epsilon_{\rm sc}$; other 
parameters are given in the corresponding panels. 
{\it Panels a and d}:  $\epsilon_{\rm sc,z}=0.03$ with
$i_{\rm in, out,0}=75^\circ$, and $i_{\rm in,z,0}=5^\circ$
(setting  $i_{\rm out,z,0}=5^\circ$ and 
$\Omega_{\rm out,z}=180^\circ$). 
The binary is nearly unaffected by the axisymmetric cluster and
it undergoes LK oscillations with small perturbations due to small changes of
$j_{\rm z}$,  reaching $e_{\rm max}\simeq(1-5/3\cos^2 i_{\rm in, out,0})^{1/2}$.
{\it Panels b and e}:  $\epsilon_{\rm sc,z}=1$ with
$i_{\rm in, out,0}=15^\circ$, $i_{\rm out,z,0}=50^\circ$, and 
$\Omega_{\rm out,z,0}=0$.
The orbit evolution is chaotic and undergoes a random walk through 
most of the  allowed phase space (this behavior is 
also depicted in panel d of Figure \ref{fig:section}).
Even though $i_{\rm in, out}$ is initially small for LK oscillations
to occur, the binary flips to retrograde inclinations and it reaches
$1-e\sim10^{-4}$ after $\sim5$ cycles.
{\it Panels c and f}:  $\epsilon_{\rm sc,z}=50$ with
$i_{\rm in, out,0}=45^\circ$ and $i_{\rm in, z,0}=80^\circ$
(by setting $i_{\rm out,z,0}=35^\circ$ and $\Omega_{\rm out,z,0}=0$).
The binary maximum 
eccentricity is set by its initial mutual inclination with respect to the cluster 
 symmetry axis  $\hat{\bf n}_z$: $e_{\rm max}=(1-5/3\cos^2 i_{\rm in,z,0})^{1/2}=0.974$.
Note that the inclination (and eccentricity) oscillations have a period of 
$\sim10\tau_{ \mbox{\tiny LK}}$, consistent
with Equation (\ref{eq:tau_kl_z}).}
\label{fig:exps}
\end{figure*}

 \paragraph{Numerical example}

In Figure \ref{fig:example1} we show an example of the orbital
evolution of a binary integrating the full model (Eqs. [\ref{eq:edot}]-[\ref{eq:jout_evol}])
for a case with $\epsilon_{\rm sc,z}=0.3$ (black lines).

From panel a we observe that the eccentricities reach values
of $1-e\sim10^{-5}$ compared to $e_{\rm max}=\sqrt{1-5/3\cos^2 (72^\circ)}=0.917$
for the case with $\epsilon_{\rm sc,z}=0$ (red line).
The binary flips its orbit relative to the outer orbit from prograde to 
retrograde and back during the large eccentricity periods
(panel b).
This behavior resembles the orbit flips in the three-body interactions
with octupole-level perturbations (e.g., \citealt{katz11,LN11}) 
and are due to the forcing of $j_{\rm z}$,
which periodically changes sign (panel c).
Given the parameters of this example (initial $i_{\rm out,z}=70^\circ$ and
$\epsilon_{\rm sc,z}=0.3$), from Equation (\ref{eq:jz_evol}) we have 
that $d j_z/d\tau\sim0.1\sin i\sin\Omega$, meaning that $j_{\rm z}$
changes by $\lesssim0.1$ in every LK cycle, consistent 
with small-amplitude oscillations observed in panel c.
The argument of periapsis librates initially at around $90^\circ$ (panel d)
and quasi-periodically switches to libration around  $-90^\circ$ and back.

In conclusion, this first example shows that the effect of having nodal
precession of the outer orbit can be understood (at least in 
this specific regime)
as having an external perturbation in the three-body interaction 
that forces $j_{\rm z}$.

\begin{figure*}\centering 
\includegraphics[angle=0,width=17.5cm]{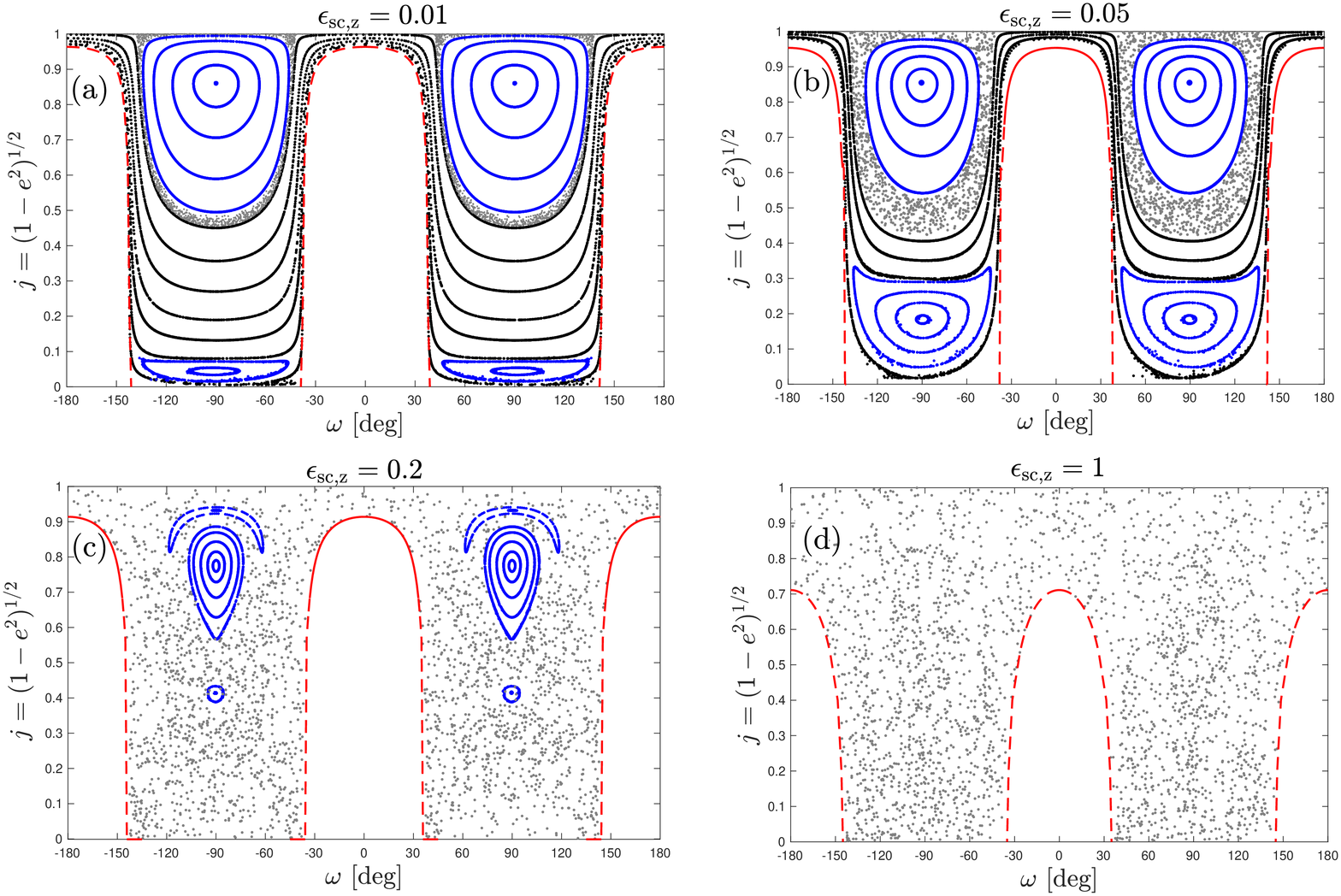}
\caption{Surfaces of section in the $\omega-j$ space 
of the two-degree of freedom 
Hamiltonian $\tilde{\mathcal{H}}_{\rm rot}$ in Equation (\ref{eq:H_BHs}) obtained by 
solving the equations of
motion (\ref{eq:edot_2})-(\ref{eq:jdot_2}) for different values
of $\epsilon_{\rm sc,z}$ as labeled. 
The blue and black lines show the librating and circulating trajectories, respectively, while
the gray dots indicate the chaotic regions and the minimum $j=\sqrt{1-e^2}$ allowed
by conservation of $\tilde{\mathcal{H}}_{\rm rot}$ is shown in red.
Each point corresponds to one crossing of the 
$\Omega=0$ plane with positive direction $d\Omega/dt>0$.
In all panels we have fixed the energy to
$\tilde{\mathcal{H}}_{\rm rot}=0.1$ with $i_{\rm out,z}'=45^\circ$
and $\Omega_{\rm out,z}'=0$.
The red lines correspond to the maximum eccentricities
allowed by the conservation of $\tilde{\mathcal{H}}_{\rm rot}$.
{\it Panel a: } $\epsilon_{\rm sc,z}=0.01$. 
The trajectories are similar to the integrable LK Hamiltonian but the small
pertubations due to the cluster introduces a narrow chaotic region
around the separatrix and a new low$-j$ libration region.
{\it Panel b: } $\epsilon_{\rm sc,z}=0.05$. The chaotic regions around 
the separatrix expand and the low$-j$ separatrix expands reaching up 
to $j\sim0.3$.
{\it Pamel c:} $\epsilon_{\rm sc,z}=0.2$.  
The two libration regions overlap and gives rise to widespread 
chaos around them, while a new high$-j$ libration islands appears.
{\it Panel d: } $\epsilon_{\rm sc,z}=1$. The libration islands disappear
and the chaotic region occupies all the available phase-space.
}
\label{fig:section}
\end{figure*}

\subsection{Dynamical behavior for 
different values of $\epsilon_{\rm sc,z}$}

In this section, we discuss the dynamical behavior for different 
values of $\epsilon_{\rm sc,z}$ and use Figures \ref{fig:exps}
and \ref{fig:section} to illustrate the general trends.

In Figure \ref{fig:exps} we show a set of characteristic examples
of the evolution of the binary for small, intermediate, and large values 
of $\epsilon_{\rm sc,z}$ (left to right panels, respectively). We
integrate the Equations of motion from the full model 
(Eqs. [\ref{eq:edot}]-[\ref{eq:jout_evol}]) with different 
initial conditions (see the corresponding labels).

In Figure \ref{fig:section} we show the surfaces of section
of the two-degree of freedom Hamiltonian $\tilde{\mathcal{H}}_{\rm rot}$ in Equation
(\ref{eq:H_BHs}) in $\omega-(1-e^2)^{1/2}$ space to show how the
increase in the perturbation strength parametrized by $\epsilon_{\rm sc,z}$ 
leads to the overlap between the LK resonance and a new resonance 
that appears at lower $j$.
This resonance overlap, and possibly others, gives rise to the widespread chaos 
observed as $\epsilon_{\rm sc,z}$ approaches unity.
We remark that a detailed study of the chaotic behavior 
in our model is beyond the scope of our work 
as $\tilde{\mathcal{H}}_{\rm rot}$ depends on two parameters
($\epsilon_{\rm sc,z}$ and $i_{\rm out}$) and sections for various 
energy manifolds should be considered.
Thus, this figure should only be considered  as a
representative example.

To create the surfaces of section, we evolve the binaries using the equivalent 
potential in the rotating frame from Equation (\ref{eq:phi}) for which the 
equations of motion can be written as
\ba
 \frac{d {\bf e}}{d\tau}&=& 2\jvec\times\evec
 -5\left(\evec\cdot\hat{\bf j}_{\rm out}\right)\jvec\times\hat{\bf j}_{\rm out}+
\left(\jvec\cdot\hat{\bf j}_{\rm out}\right)\evec\times\hat{\bf j}_{\rm out}\nonumber\\
&&-\epsilon_{\rm sc,z}\left(\hat{\bf n}_z\cdot \hat{\bf j}_{\rm out}\right)
\evec\times\hat{\bf n}_z  
\label{eq:edot_2}\\
\frac{d {\bf j}}{d\tau}&=&
 \left({\bf j}\cdot\hat{\bf j}_{\rm out}\right){\bf j} \times \hat{\bf j}_{\rm out}
-5\left({\bf e}\cdot\hat{\bf j}_{\rm out}\right){\bf e}\times\hat{\bf j}_{\rm out}
\nonumber\\
&&-\epsilon_{\rm sc,z}\left(\hat{\bf n}_z\cdot \hat{\bf j}_{\rm out}\right)
\jvec\times\hat{\bf n}_z,
\label{eq:jdot_2}
\ea
and vary $\epsilon_{\rm sc,z}$, while fixing 
$\tilde{\mathcal{H}}_{\rm rot}=0.1$,  $i_{\rm out,z}'=45^\circ$,
and $\Omega_{\rm out,z}'=0$.

\subsubsection{Limit of small  $\epsilon_{\rm sc,z}$: 
slow nodal precession}

For $\epsilon_{\rm sc,z}\ll1$, the Hamiltonian in Equation
(\ref{eq:H_BHs}) roughly coincides with the integrable 
Lidov-Kozai Hamiltonian.
Thus, we expect that the binaries starting from low-eccentricity
orbits and mutual inclinations $i_{\rm in,out,0}=\cos^{-1}\hat{\bf {j}}_0\cdot \hat{\bf j}_{\rm out}$
reach a maximum eccentricity roughly given by
\ba
e_{\rm max}=\left(1-\frac{5}{3}\cos^2 i_{\rm in,out,0}\right)^{1/2}.
\label{eq:emax_kl}
\ea

In the left panels of Figure \ref{fig:exps} we show the evolution for a binary
with $\epsilon_{\rm sc,z}=0.03$ and find that indeed  the maximum
eccentricity coincides with this expression (panel d).
As expected, the evolution of the mutual inclinations  $i_{\rm in,out}$
follow the regular LK oscillations, while the evolution of the inclination
relative to $|\hat{\bf n}_z|$, $i_{\rm in,z}$, follows from the nodal 
precession of the binary
 around $\hat{\bf j}_{\rm out}$ and the LK inclination oscillations.

An exception to the evolution depicted above occurs when $|j_{\rm z}|$ 
is small, $|j_{\rm z}|\lesssim\epsilon_{\rm sc,z}$.
From Equation (\ref{eq:jz_evol}), we estimate that for small
$|j_{\rm z}|$ ($|\sin i_0|\sim1$) the changes in $j_{\rm z}$
over one LK cycle are
\ba
|\Delta j_{\rm z}|\sim\frac{1}{2}\epsilon_{\rm sc,z}|\sin 2i_{\rm out,z} |,
\label{eq:delta_j}
\ea
and, therefore, the cluster potential can drive the systems with 
$|j_{\rm z}|\lesssim\epsilon_{\rm sc,z}$ to cross $j_{\rm z}=0$.
By crossing $j_{\rm z}=0$, the eccentricities can greatly overcome
$e_{\rm max}$ in Equation (\ref{eq:emax_kl}). We will come back
to this point below,  in the population synthesis study of \S\ref{sec:pop_eps}
(see yellow line in Figure \ref{fig:population_epsz}).

In turn, the surface of section in Figure \ref{fig:section} with $\epsilon_{\rm sc,z}=0.01$
(panel a) shows that the trajectories are similar to those from the well-known
integrable Lidov-Kozai Hamiltonian with the librating paths for low $j$
at around the fixed point at $|\omega|=90^\circ$ and circulating trajectories 
outside this region. 
The small perturbation due to the cluster potential
introduces two main changes.
First, the trajectories around the separatrix become chaotic (gray dots)
similar to the ocupole-level pertubations in the LK mechanism 
seen by \citet{holman97} with similar perturbation strengths  
with $\epsilon_{\rm oct}\sim0.008$ in Equation
(\ref{eq:oct}).
Second, a new libration region appears at $j\lesssim0.1$ 
with a fixed point at $|\omega|=90^\circ$ and $j\sim0.03$.
From the Equation (\ref{eq:edot_2}), we observe that a fixed
point ($d {\bf e}/d\tau=0$) involving the term due to the cluster
$\epsilon_{\rm sc,z}\ll1$  have to occur for
$j\sim\epsilon_{\rm sc,z}$, explaining why the resonance 
appears at such small values.
As we discuss next, this new resonance plays a major role 
in the evolution of the system as it expands with increasing
$\epsilon_{\rm sc,z}$ and leads to global chaos.

Some intuition regarding the chaos around the separatrix might 
be gained by looking at the dynamically equivalent and time-dependent
one-degree-of-freedom Hamiltonian in Equation (\ref{eq:phi_t}),
where $\hat{\bf j}_{\rm out}(t)$ changes steadily due to the nodal
precession (Eq. [\ref{eq:jout_t}]).
Here, the variation of $\hat{\bf j}_{\rm out}(t)$ changes the location 
of the separatrix in time, while repeated separatrix crossings 
can act to drive stochasticity \citep{LL83}.
Thus, chaos might be expected to occur only for the regions in
phase-space that are close enough to the separatrix and
such distance depends on the strength of the perturbation.

In panel b, we show the surface of section for 
$\epsilon_{\rm sc,z}=0.05$ and observe that the phase-space
structure is similar to the case with  $\epsilon_{\rm sc,z}=0.01$
and the two libration islands remain separated in action-space.
The main effect of increasing $\epsilon_{\rm sc,z}$  is to
increase the area of the low$-j$ separatrix, whose respective 
fixed point moves from $j\sim0.05$ for $\epsilon_{\rm sc,z}=0.01$
to  $j\sim0.2$ for $\epsilon_{\rm sc,z}=0.05$. 
Also, the area covered by the chaotic region around the high$-j$, or
 Lidov-Kozai,
sepatrarix increases by a factor of a few and reaches down to 
$j\sim 0.4$.
As a result, the distance between the chaotic region and the separatrix
of the low$-j$ resonance decreases from $\Delta j \sim0.4$ for 
 $\epsilon_{\rm sc,z}=0.01$ 
to  $\Delta j\sim0.1$ for $\epsilon_{\rm sc,z}=0.05$.
As we increase $\epsilon_{\rm sc,z}$ this distance approaches
zero and the resonance overlaps leading to 
widespread chaos as observed in panels c and d.
We discuss this overlapped regime next.

Finally, we remark that the chaos in the small
$\epsilon_{\rm sc,z}$ regime occurs only around the separatrix
and does not promote the excitation of large eccentricities.
The excitation to extreme eccentricities happen 
for orbits with $|j_{\rm z}|\lesssim\epsilon_{\rm sc,z}$, in a
similar fashion as the octupole-level perturbations of the
LK mechanism that drives $j_{\rm z}$ towards 0.

\subsubsection{Intermediate values of $\epsilon_{\rm sc,z}$}

For intermediate values of $\epsilon_{\rm sc,z}$ the
behavior is more complicated than the  limiting
cases of small and large $\epsilon_{\rm sc,z}$.
In particular, the dynamical evolution of the binary
can be governed by chaotic behavior in a large 
fraction of the available phase-space.

In the middle panels of Figure \ref{fig:exps} we show the evolution for a 
binary with $\epsilon_{\rm sc,z}=1$ and observe that
the inclinations behave in a chaotic-looking fashion (panel b).
Most remarkably, starting from a low mutual inclination,
$i_{\rm in,out}=15^\circ$, the binary orbit becomes retrograde  after
$\sim5$ cycles.
In turn, the evolution in $e-\omega$ space (panel e) shows 
rapid intermittence between librating and circulating trajectories,
typical of chaotic systems, and the eccentricity reaches
 $1-e\sim 10^{-4}$ after $\sim5$ cycles.

In Figure \ref{fig:section} we show the surfaces of section
for $\epsilon_{\rm sc,z}=0.2$ (panel c) and $\epsilon_{\rm sc,z}=1$ (panel d)
 in which wide chaotic regions can be observed.
 
For the case with $\epsilon_{\rm sc,z}=0.2$ we observe 
that there exists three libration islands separated in action space.
Presumably, the two rounder libration regions  with corresponding fixed points 
at $j\sim0.75$ and  $j\sim0.4$ are the same libration regions observed
in the lower $\epsilon_{\rm sc,z}$ cases (panels a and b), but have significantly
shrunk in area because these resonances overlap and they are well-embedded
in a chaotic sea. 
The banana-shaped libration island at higher $j$ is a new resonance
and, admittedly, there can be other very narrow resonances 
that are not captured by our set of initial conditions.
Most interestingly, we observe that outside these libration islands, which correspond
to a small area of the available phase-space indicated by the red lines,
the motion is chaotic. Thus, most initial conditions
can diffuse towards low angular momentum: 
$j\to 0$ or, equivalently,  $e\to1$.

As we increase $\epsilon_{\rm sc,z}$ from 0.2 to 1 (panel d),
we observe that chaos is widespread and it reaches all the 
available phase-space. 
Thus, all initial conditions reach
$e\to1$ given enough time for the stochastic diffusion  
to populate the low$-j$ regions of phase-space.
We come back to this issue of the diffusion timescale in section
\S\ref{sec:Nmax}, where we show in Figure \ref{fig:population_Nmax}
that the fraction of low$-j$ orbits depends smoothly on the evolution
 time.

Finally, we remark that widespread chaos is not a condition to reach $1-e\ll1$
and quasi-periodic orbits can reach extreme eccentricities 
simply by crossing $j_{\rm z}=0$, similar to the behavior 
described in the previous section for small 
$\epsilon_{\rm sc,z}$. The main difference with the small 
$\epsilon_{\rm sc,z}$ case is that such crossing can occur
for initially larger volume of the phase-space since a rough
necessary  condition is to have $|j_{\rm z}|\lesssim\epsilon_{\rm sc,z}$.

\subsubsection{Limit of large $\epsilon_{\rm sc,z}$: 
fast nodal precession}
\label{sec:large_eps}

In the limit of large $\epsilon_{\rm sc,z}$
the binary outer orbit precesses around $\hat{\bf n}_z$
on a timescale that is much shorter than  $\tau_{ \mbox{\tiny LK}}$.
We can express the potential in Equation (\ref{eq:phi})
in the reference frame of the cluster principal  axes 
\ba
\frac{\bar{\Phi}}{\phi_0}=
\frac{1}{6}-e^2 -\frac{1}{2}[{\bf {j}}\cdot \hat{\bf j}_{\rm out}(t)]^2+
\frac{5}{2}  [{\bf e}\cdot \hat{\bf j}_{\rm out}(t)]^2
\label{eq:phi_t}
\ea
with 
\ba  
\hat{\bf j}_{\rm out}(t)&=&\cos \Omega_{\rm out,z}(t) \sin i_{\rm out,z}\hat{{\bf n}}_x+
\sin \Omega_{\rm out,z}(t) \sin i_{\rm out,z} \hat{{\bf n}}_y\nonumber \\
&&+\cos i_{\rm out,z}\hat{{\bf n}}_z,
\label{eq:jout_t}
\ea
and average this potential over the fast varying angle 
$\Omega_{\rm out,z}(t)\propto t$ (i.e., keeping {\bf {e}} and ${\bf {j}}$ constant in one 
precession cycle) as\footnote{An equivalent expression is easily found using the Hamiltonian 
written in terms of orbital elements (see, e.g., Equation 20 in \citealt{naoz13}) 
after averaging over the fast varying longitude of the ascending 
nodes.}
\ba
\frac{\left<\bar{\Phi}\right>_{\Omega}}{\phi_0}&=&\frac{1}{2\pi}\int d\Omega_{\rm out,z}
\frac{\bar{\phi}}{\phi_0}=\left(1-\frac{3\sin^2 i_{\rm out,z}}{2}\right)\nonumber\\
&&\times
\left[\frac{1}{6}-e^2 -
\frac{1}{2}({\bf {j}}\cdot \hat{\bf n}_z)^2+
\frac{5}{2}  ({\bf e}\cdot \hat{\bf n}_z)^2\right].
\label{eq:phi_z}
\ea
Remarkably, this potential coincides with the original one in Equation 
(\ref{eq:phi_t}) but multiplied by the constant term $(1-3/2\sin^2 i_{\rm out,z})$
and replacing $\hat{\bf j}_{\rm out}$ by $\hat{{\bf n}}_z$.
This potential is
axisymmetric with respect to $\hat{\bf n}_z$ and, therefore, 
$\ell_{\rm z}\equiv\sqrt{1-e^2}\cos i_{\rm in, z}$ with
 $i_{\rm in,z}=\cos^{-1}\hat{\bf {j}}\cdot \hat{\bf n}_z$
is a constant of motion.
Thus, from this potential we can determine the maximum eccentricity, which is reached
at $\omega=\pi/2$ and $3\pi/2$ (i.e., $[\hat{\bf e}\cdot \hat{\bf n}_z]^2=\sin^2 i$),  
starting from $e=0$ and inclination $i_{\rm z,0}$ as
\ba
e_{\rm max}=\left(1-\frac{5}{3}\cos^2 i_{\rm in, z,0}\right)^{1/2},
\label{eq:emax_z}
\ea
meaning that  we recover the classic expression
of the maximum eccentricity from the LK mechanism but 
with the inclination of the inner orbit relative to $\hat{\bf n}_z$,
instead of $\hat{\bf j}_{\rm out}$. 

The precession amplitude, which 
is determined by  $\sin i_{\rm out,z}$, only changes the magnitude of
the potential and, therefore, these new LK eccentricity oscillations 
occur in a timescale
\ba
\tau_{ \mbox{\tiny LK}}|_{\epsilon_{\rm sc,z}\to\infty}=\left|1-\frac{3\sin^2 i_{\rm out,z}}{2}\right|^{-1}
\tau_{ \mbox{\tiny LK}},
\label{eq:tau_kl_z}
\ea
where $\tau_{ \mbox{\tiny LK}}$ is defined in Equation 
(\ref{eq:tau_lk}).
We have checked numerically that this expression for the eccentricity oscillations
scales well with different values of $ i_{\rm out,z}$. 
In particular, we checked that the timescale of these oscillations indeed
diverges as $\sin^2 i_{\rm out,z}\to2/3$.

In the panels c and f of Figure \ref{fig:exps} we show an example of the 
evolution for $\epsilon_{\rm sc,z}=50$ and find that, as explained above,
$i_{\rm in,z}$ (red curve) shows regular LK cycles, while
the inclination $i_{\rm in,out}$ (black line) 
oscillates rapidly around the $i_{\rm in,z}$.
The maximum eccentricity is given 
roughly by $e_{\rm max}$ in Equation (\ref{eq:emax_z}) and 
we checked that this expression is exact for much larger 
$\epsilon_{\rm sc,z}$ (not shown).
Also, we observe that the period of  the oscillation (red curve in panel c)
is $\sim10\tau_{ \mbox{\tiny LK}}$, consistent with the 
longer period due to the extra multiplicative term of
$|1-3/2\sin^2 i_{\rm out,z}|^{-1}\simeq8.35$
in Equation (\ref{eq:tau_kl_z}).
As expected from the potential in Equation (\ref{eq:phi_z})
one recovers the standard LK dynamics, but now the orbital 
elements have to be defined relative to the cluster orientation
(see, e.g., the circulating trajectory in $e-\omega_{\rm in,z}$ from
panel f).

In summary, for large $\epsilon_{\rm sc,z}$ the cluster induces
fast nodal precession so the time-averaged potential of the outer 
binary acting in the inner one is axi-symmetric around 
$\hat{\bf n}_z$ and, mathematically equivalent to that 
of the LK mechanism.
Thus, this regime can only excite extreme eccentricities
for binaries with $i_{\rm in,z}\sim90^\circ$.

\begin{figure}
\includegraphics[,width=8.5cm]{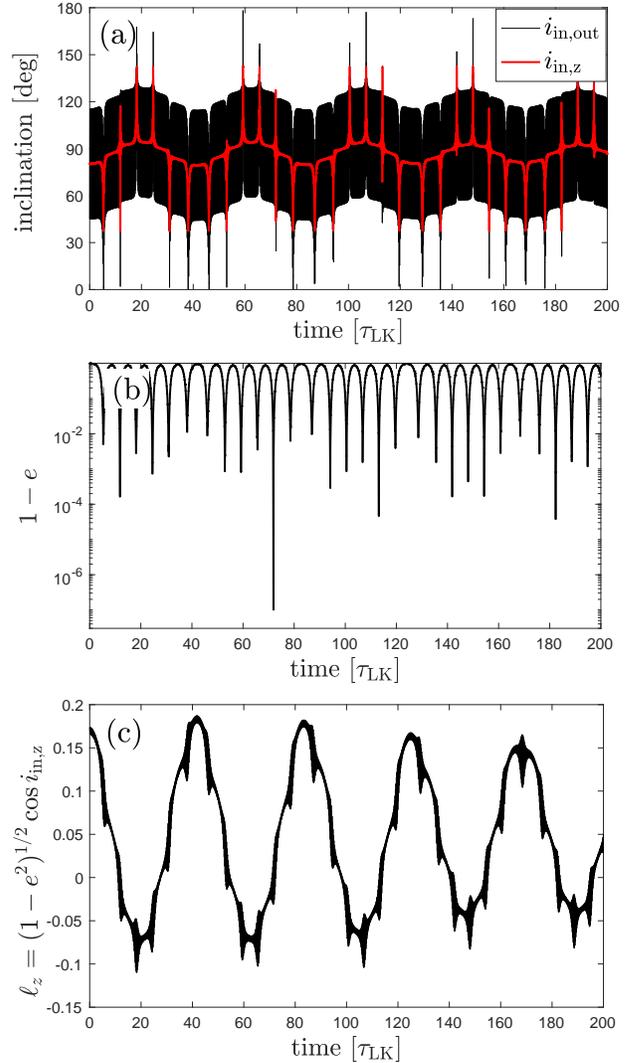}
\caption{Evolution of a binary in a fully triaxial cluster with 
$\epsilon_y/\epsilon_z=0.2$  and  $\epsilon_{\rm sc,z}=50$.
This figure is similar to the example  in
panels c and f of Figure \ref{fig:exps} for comparison with the
axisymmetric case ($\epsilon_y=0$).
The initial conditions are $\omega=0$, $\Omega=0$, $e_{\rm in}=e_{\rm out}=0.01$,
$i_{\rm in, out}=45^\circ$,  and $i_{\rm in, z,0}=80^\circ$
(by setting $i_{\rm out,z,0}=35^\circ$ and $\Omega_{\rm out,z,0}=90^\circ$).
{\it Panel a:} inclination relative to $\hat{{\bf n}}_z$, $i_{\rm in,z}$,
and the inclination relative to outer orbit $i_{\rm in,out}$, which varies rapidly 
because ${\rm j}_{\rm out}$ is rapidly precessing. The orbit flips due to 
the extra precession source around $\hat{{\bf n}}_y$.
{\it Panel b:} the eccentricity reaches $1-e_{\rm max}\lesssim10^{-7}$
compared to $e_{\rm max}=\sqrt{1-5/3\cos^2 i_{\rm in,z,0}}=0.974$ in the case
with  $\epsilon_y=0$.
Note that the timescale of these oscillations is $\sim10\tau_{ \mbox{\tiny LK}}$
because of the extra multiplicative term of
$|1-3/2\sin^2 i_{\rm out,z}|^{-1}\simeq 8.35$
in Equation (\ref{eq:tau_kl_z}).
{\it Panel c:} specific angular momentum $\ell_{\rm z}={\bf j}\cdot\hat{{\bf n}}_z$.
It changes because $\epsilon_y>0$ and extreme eccentricities occur when 
$j_{\rm z}=0$.
}
\label{fig:eps_y}
\end{figure}

\begin{figure*}[t!]
\center
\includegraphics[width=18cm]{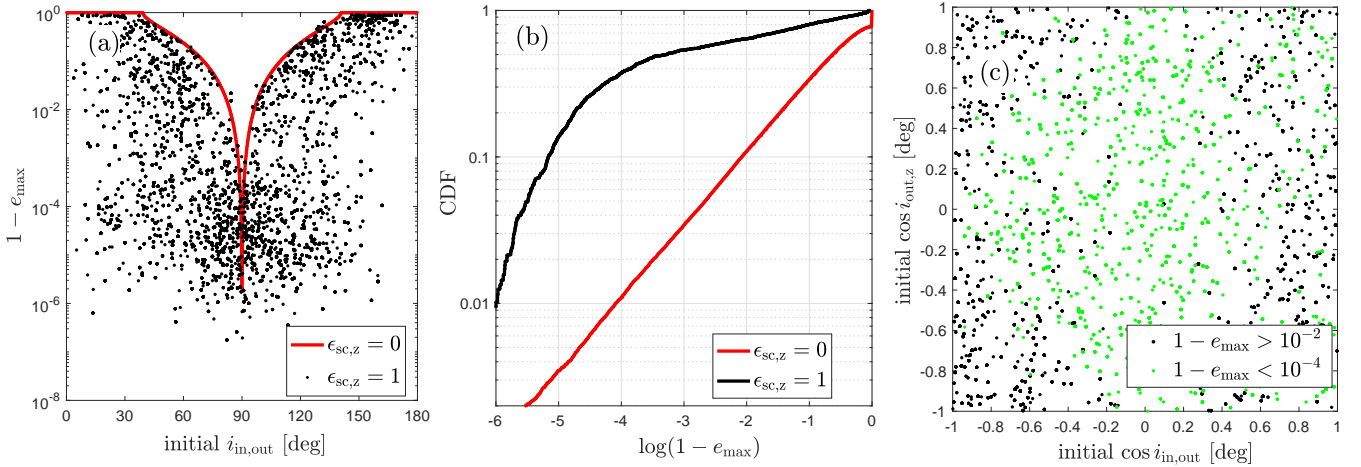}
\caption{Maximum eccentricities reached after 100 Lidov-Kozai
cycles for 10,000 binaries  embedded
in a cluster with $\epsilon_{\rm sc,z}=1$ (Eq. [\ref{eq:omega_sz}]).
The initial conditions for the binaries are $e_{\rm in}=e_{\rm out}=0.1$,
and random isotropic vectors $\hat{\bf j}_{\rm in}$ and 
$\hat{\bf j}_{\rm out}$.
{\it Panel a:} maximum eccentricity as a function of 
$i_{\rm in,out}$ (angle between
$\hat{\bf j}_{\rm in}$ and $\hat{\bf j}_{\rm out}$).
The solid lines correspond to the case without a cluster ($\epsilon_{\rm sc,z}=0$),
which results in $e_{\rm max}=\sqrt{1-5/3\cos^2i_{\rm in,out}}$.
{\it Panel b:} cumulative distribution of the maximum eccentricities.
{\it Panel c:}  initial orientations $i_{\rm in,out}$  and
 $i_{\rm z,out}$ (angle between
$\hat{\bf j}_{\rm out}$ and $\hat{\bf n}_{\rm z}$) for
two samples: systems that reach  $1-e_{\rm max}<10^{-4}$ 
(green dots) and systems that do not (black dots).}
\label{fig:population}
\end{figure*}

\subsection{Effect from triaxiality}
\label{sec:epsy}

We briefly discussed the main effects from including a triaxial
component to stellar 
cluster, meaning that $0<\epsilon_{\rm y}<\epsilon_{\rm z}$.

We expect that in the regime of small $\epsilon_{\rm sc,z}$, adding an extra
source of slow nodal precession around $\hat{\bf n}_y$ will not change the 
behavior relative to the axisymmetric case and the maximum eccentricity 
will be given by the classic LK expression.
On the contrary, as we will show, for large  $\epsilon_{\rm sc,z}$ adding a slower
precession around $\hat{\bf n}_y$ on top of the dominant precession 
around $\hat{\bf n}_z$ can have important effects.
We ignore the complicated regime in which 
$\epsilon_{\rm sc,y}\lesssim\epsilon_{\rm sc,z}\sim1$ because there 
are too many parameters that might change the phase-space structure 
($\epsilon_{\rm sc,y}$, $\epsilon_{\rm sc,z}$ and orientations 
$\Omega_{\rm out,z}$, $ i_{\rm out,z}$) and leave it for a future study.

In Figure \ref{fig:eps_y} we show an example in the large 
 $\epsilon_{\rm sc,z}$ regime, similar to that in panels 
 c and f of Figure \ref{fig:exps}, which we discussed in \S\ref{sec:large_eps}.
 The only difference is that we set $\epsilon_y=10$, so $\epsilon_y/\epsilon_z=0.2$.
 This means that the time evolution of $\hat{\bf j}_{\rm out}$ is 
 described by the
following equations (Eq. (\ref{eq:jout_evol}) with $1-e_{\rm out}\ll1$):
\ba
\frac{d \hat{\bf j}_{\rm out}}{d\tau}&=&\epsilon_{\rm sc,z}\bigg[ 
\left(\hat{\bf n}_z\cdot \hat{\bf j}_{\rm out}\right)\hat{\bf j}_{\rm out}\times
 \hat{\bf n}_z+\nonumber\\
&&+\frac{\epsilon_{\rm y}}{\epsilon_{\rm z}}
 \left(\hat{\bf n}_y\cdot \hat{\bf j}_{\rm out}\right)\hat{\bf j}_{\rm out}\times
 \hat{\bf n}_y
 \bigg],
\ea
which describes the precession of $\hat{\bf j}_{\rm out}$
around  both $\hat{\bf n}_y$ and $\hat{\bf n}_z$.
We observe that adding this slower precessional frequency around
$\hat{\bf n}_y$ creates a extra modulation in the inclinations (panel a) compared to the
regular motion in panel c of Figure \ref{fig:exps}, and the orbit flips
every $\sim4$ oscillations, roughly coincident with the ratio of precession timescales 
$\epsilon_{\rm z}/\epsilon_{\rm y}=5$.
This longer modulation is better observed in panel c 
where we show the evolution of $\ell_z$, which is constant
for $\epsilon_y=0$, and observe that in this example it can cross 
$\ell_z=0$. 
During these crossings, the eccentricity reaches extreme values (panel b)
and reaches up to $1-e\sim10^{-7}$ after a time of $\sim70\tau_{\mbox{\tiny LK}}$.
Recall that in this particular example, the oscillations take extra time 
compared to other choices because of the dilation term
$|1-3/2\sin^2 i_{\rm out,z}|^{-1}\simeq 8.35$
in Equation (\ref{eq:tau_kl_z}). Other choices of $i_{\rm out}$ can
make the timescales of the oscillations comparable to 
$\tau_{\mbox{\tiny LK}}$.

In order to have an intuitive understanding of the dynamics
 we give an heuristic argument based on the previous section
 and the assumption that the level of triaxiality is small,
$\epsilon_{\rm y}/\epsilon_{\rm z}\ll1$.
Thus, we can treat this new precession source 
as a perturbation to the simple 
axisymmetric potential in Equation (\ref{eq:phi_z}).
In this regime, we expect that averaging the LK potential over the motion
of $\hat{\bf j}_{\rm out}(t)$, as we did in \S\ref{sec:large_eps},
creates a strong axisymmetric potential in the direction of  $\hat{\bf n}_z$
with a small ($\mathcal{O}[\epsilon_{\rm y}/\epsilon_{\rm z}]$) component 
in the $\hat{\bf n}_y$ direction.
If so, this means that the resulting potential should have a
terms $\propto \epsilon_{\rm y}/\epsilon_{\rm z} ({\bf {j}}\cdot \hat{\bf n}_y)
({\bf {j}}\cdot \hat{\bf n}_z)$ plus other terms involving ${\bf {e}}$.
Thus, $\ell_z={\bf {j}}\cdot \hat{\bf n}_z$ is no longer fixed and 
would change as $\dot{\ell}_z  \propto\epsilon_{\rm y}/\epsilon_{\rm z} \sin i_{\rm out,z} 
\ell_z$, which would explain the longer period oscillations
of $\ell_z$ observed in panel c.

In summary, adding the triaxial component to the model affects
most dramatically  the regime of fast nodal precession ($\epsilon_{\rm sc,z}\gg1$) 
because $\ell_z$ can change slowly in time and cross $0$, allowing
the eccentricity to reach extremely high values.


\section{Population synthesis}
\label{sec:pop_synth}

We explore the effect from the cluster tidal field
for a large population of binaries and how it depends on various parameters. 
We focus on the statistical distributions of maximum 
eccentricities ($e_{\rm max}$) because 
in the secular approximation its evolution
mainly determines whether a merger is possible or not.

With the exception of our results 
in \S\ref{sec:eps_GR} we ignore the effect from relativistic 
precession in our calculations.

\subsection{Fiducial case with $\epsilon_{\rm sc,z}=1$}

In Figure \ref{fig:population}, we show the maximum eccentricities for randomly 
oriented binaries in a cluster with $\epsilon_{\rm sc,z}=1$ ($\epsilon_{\rm sc}=10$
and $\epsilon_{\rm z}=0.1$ in Eq. [\ref{eq:omega_sz}]).
We set  $e_{\rm in}=0.1$ initially so the excitation of eccentricities is 
clearly visible in the plots, while at the same time it avoids extremely small 
$e_{\rm in}$ that lie near the separatrix,
leading to long-timescale LK cycles and
possibly to enhanced chaotic behavior.

In panel a, we show how the maximum eccentricities depend on 
$i_{\rm in,out}$ (angle between
$\hat{\bf j}_{\rm in}$ and $\hat{\bf j}_{\rm out}$) and compare these with 
the expression  $e_{\rm max}=\sqrt{1-5/3\cos^2i_{\rm in,out,0}}$ 
applicable in the absence of the cluster potential 
($\epsilon_{\rm sc,z}=0$).
Probably not surprisingly, we observe that most binaries
reach $1-e_{\rm max}\ll1-\sqrt{1-5/3\cos^2i_{\rm in,out,0}}$ (most black
dots are well below the solid red line).
The distribution of  $1-e_{\rm max}$ shown in panel b depicts more clearly
this behavior and we observe that the case with $\epsilon_{\rm sc,z}=1$ 
has a much shallower cumulative distribution towards smaller $1-e_{\rm max}$ 
compared to the runs with  $\epsilon_{\rm sc,z}=0$. 
In particular, $\sim50\%$ ($\sim30\%$) of the binaries
in the runs with $\epsilon_{\rm sc,z}=1$  reach 
$1-e_{\rm max}\lesssim10^{-3}$ ($1-e_{\rm max}\lesssim10^{-4}$)
compared to $\sim3\%$ ($\sim1\%$)
in the absence of the cluster potential ($\epsilon_{\rm sc,z}=0$).

In panel c we show the initial inclinations for two samples of our fiducial
run: binaries that reach  extreme eccentricities $1-e_{\rm max}<10^{-4}$ 
(green dots) and systems with $1-e_{\rm max}>10^{-2}$
(black dots).
We observe that extreme eccentricities are almost always reached
when the initial $\cos i_{\rm in,out}\sim0$, regardless of the value of
$i_{\rm out,z}$ (inclination  of the outer orbit relative to $\hat{\bf n}_z$).
This behavior is expected because these systems start with very small
$j_z$ and from Equation (\ref{eq:jz_evol}) we expect that 
the forcing due to the cluster almost always allows $j_z$ to cross 0
at which point the eccentricity can be highly excited.
On the contrary, if  the initial $|j_z|$ is closer to unity ($|\cos i_{\rm in,out}|\sim1$),
then there is a more limited set of initial conditions that allow
for extreme eccentricity excitation and this reflects on the limited range of 
$\cos i_{\rm out,z}$ that reach $1-e<10^{-4}$, as observed in
the figure.

\begin{figure}
\center
\includegraphics[width=8.2cm]{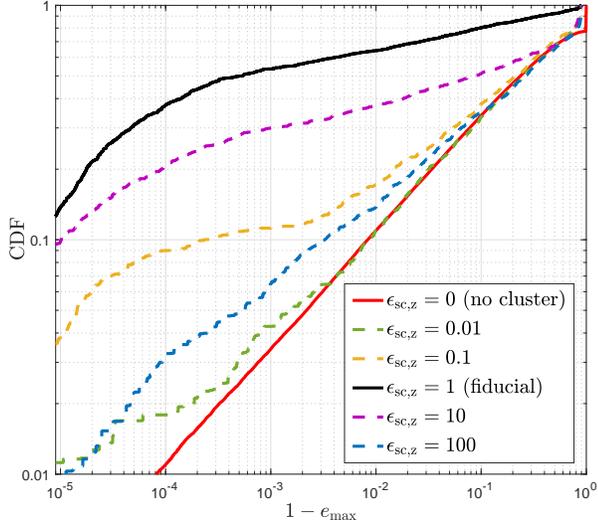}
\caption{Cumulative distribution of the maximum eccentricities 
reached after 100 Lidov-Kozai cycles for 10,000
randomly oriented  binaries  embedded
in a cluster with different values 
$\epsilon_{\rm sc,z}$ (Eq. [\ref{eq:omega_sz}]).}
\label{fig:population_epsz}
\end{figure}

\subsection{Effect from the cluster 
field strength: dependence on $\epsilon_{\rm sc,z}$}
\label{sec:pop_eps}

In Figure \ref{fig:population_epsz} we 
explore the effect of varying $\epsilon_{\rm sc,z}$.
We do so by varying $\epsilon_{\rm sc}$
and fixing $\epsilon_{\rm y}=0$ and $\epsilon_{\rm z}=0.1$
(see Equation \ref{eq:omega_sz}).
Recall that $\epsilon_{\rm sc,z}=\epsilon_{\rm z}\epsilon_{\rm sc}$
and that small values of $\epsilon_{\rm z}$ ensure that the outer orbit
remains nearly circular and only undergoes nodal precession 
around $\hat{\bf n}_z$.

Figure \ref{fig:population_epsz} gives the distributions of 
$e_{\rm max}$
for different values of $\epsilon_{\rm sc,z}$.
 We observe that our fiducial set-up
with $\epsilon_{\rm sc,z}=1$ is the most efficient at exciting extreme 
eccentricities followed by  $\epsilon_{\rm sc,z}=10$  
and $\epsilon_{\rm sc,z}=0.1$.   This behavior might be expected
because when $\epsilon_{\rm sc,z}\sim1$ the binary is subject to global 
chaos for a wide range of initial conditions, while 
for smaller values of $\epsilon_{\rm sc,z}$ the chaotic regions
shrinks significantly and might be bounded to a minimum
$j$ (see Figure \ref{fig:section}).
We expect that the chaotic region also shrinks for larger 
$\epsilon_{\rm sc,z}$ so there is a smooth transition towards 
regularity that we find for large $\epsilon_{\rm sc,z}$. 
Regardless of these differences, for all the cases above 
($\epsilon_{\rm sc,z}=\{0.1,1,10\}$) the 
excitation of extreme eccentricities, $1-e_{\rm max}\lesssim10^{-4}$,
occurs in $\sim10-50\%$ of the binaries, compared to only 
$\sim1\%$ in the absence of the cluster potential ($\epsilon_{\rm sc,z}=0$).

On the extreme ends of $\epsilon_{\rm sc,z}$ ($0.01$ or $100$ in 
Figure \ref{fig:population_epsz}) 
the behavior is more easily explained.
First, for very small values of $\epsilon_{\rm sc,z}$,  only systems with very 
small values of $j_{\rm z}$ can be appreciably affected by the cluster field. 
In particular, from Equation (\ref{eq:delta_j}) we get that the change
in $j_{\rm z}$ over one LK cycle is $\sim\epsilon_{\rm sc,z}$, meaning that
the initial conditions starting from  $|j_{\rm z,0}|\lesssim\epsilon_{\rm sc,z}$ 
can go through $j_{\rm z}=0$, unless a possible libration of $\Omega$
continues to coherently force  $j_{\rm z}$.
Since reaching $j_{\rm z}=0$ is a necessary condition to reach
extreme eccentricities and the initial distribution of $j_{\rm z}$ is uniform 
(isotropic inclinations), we expect that roughly a fraction 
$\sim \epsilon_{\rm sc,z}$ will be strongly  affected by the cluster tides.
Consistent with these expectations, we observe 
from Figure \ref{fig:population_epsz}  that 
in the $\epsilon_{\rm sc,z}=0.1$ and $\epsilon_{\rm sc,z}=0.01$ cases,
$1-e_{\rm max}$ departs from the $\epsilon_{\rm sc,z}=0$ line (no cluster case)
and flattens to to the left to reach much higher eccentricities
in $\sim10\%$ and $\sim1\%$ of the binaries (CDF of $\sim0.1$ and $\sim0.01$,
respectively).

Second, as discussed in \S\ref{sec:large_eps}, for very large  values of 
$\epsilon_{\rm sc,z}$ the maximum eccentricity is given by 
$e_{\rm max}=\sqrt{1-5/3\cos^2 i_{\rm in,z,0}}$, where 
$i_{\rm in,z,0}=\cos^{-1}\hat{\bf {j}}\cdot \hat{\bf n}_z$ initially
(Equation [\ref{eq:emax_z}]). 
Since $\hat{\bf {j}}$ is initially chosen to be isotropically 
distributed, we expect that the distributions of $e_{\rm max}$ for $\epsilon_{\rm sc,z}\gg1$
and $\epsilon_{\rm sc,z}=0$ coincide.
From Figure \ref{fig:population_epsz} we observe that the curves 
$\epsilon_{\rm sc,z}=100$ and $\epsilon_{\rm sc,z}=0$ coincide for 
$1-e_{\rm max}\gtrsim10^{-2}$, while for $1-e_{\rm max}\lesssim10^{-2}$,
$\epsilon_{\rm sc,z}=100$ is above the runs with $\epsilon_{\rm sc,z}=0$.
This departure at very large eccentricities is due to small eccentricity oscillations 
taking place in the short timescales $\tau_{\tiny \rm KL}$ when  $\epsilon_{\rm sc,z}$
is large, but finite so the approximation to get Equation (\ref{eq:emax_z}) is not 
fully fulfilled (see the wiggles around $e_{\rm max}$ in panel
f of Figure \ref{fig:exps}). 
These oscillations disappear as
we increase  $\epsilon_{\rm sc,z}$ to values $\gg100$.

\begin{figure}
\center
\includegraphics[width=8.5cm]{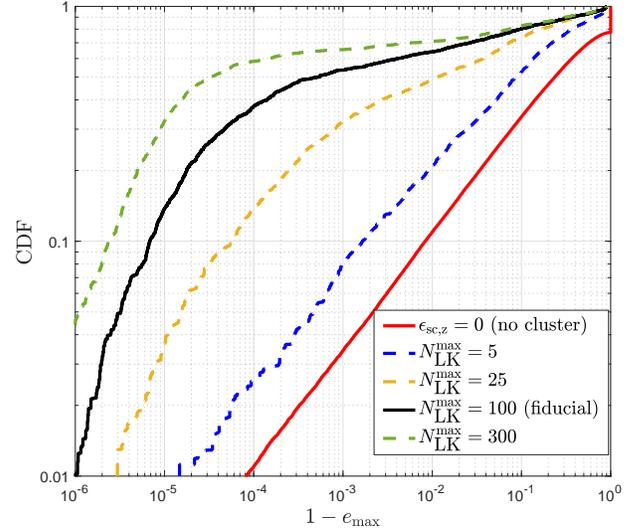}
\caption{Cumulative distribution of the maximum eccentricities 
reached after $N_{ \mbox{\tiny LK}}^{\rm max}=t_{\rm max}/\tau_{ \mbox{\tiny LK}}$
(as labeled) for 10,000 randomly oriented binaries  embedded
in a cluster with $\epsilon_{\rm sc,z}=1$. The fiducial simulation with 
$N_{ \mbox{\tiny LK}}^{\rm max}=100$ is shown in the solid black line.
The gradual increase in the number of binaries 
reaching extreme eccentricities with increasing 
$N_{ \mbox{\tiny LK}}^{\rm max}$ is expected from the 
chaotic diffusion of the eccentricities.}
\label{fig:population_Nmax}
\end{figure}

\subsection{Effect from maximum evolution time: number of LK
cycles $N_{ \mbox{\tiny LK}}^{\rm max}$}
\label{sec:Nmax}

In the cluster environment the binaries can be subject to stellar 
evolution or evaporation before a merger can happen. Thus, it is 
important to understand how the maximum evolution time affects
the merger fractions.

We give the maximum integration time in units of LK cyles
as $N_{ \mbox{\tiny LK}}^{\rm max}=t_{\rm max}/\tau_{ \mbox{\tiny LK}}$, which 
can be easily converted to absolute timescales 
using Equation (\ref{eq:tau_lk}) or compared directly to the 
collision time $N_{ \mbox{\tiny LK, coll}}$ and
evaporation time $N_{ \mbox{\tiny LK, evap}}$
in Equations (\ref{eq:Ncoll})
and (\ref{eq:Nevap}), respectively.

In Figure \ref{fig:population_Nmax} we show the distribution of
$e_{\rm max}$ for different values of  $N_{ \mbox{\tiny LK}}^{\rm max}$
from 5 to 300 and observe that the fraction of systems
reaching $1-e_{\rm max}\sim10^{-4}-10^{-5}$ increases gradually 
with $N_{ \mbox{\tiny LK}}^{\rm max}$. 
This gradual increase is expected from the chaotic 
diffusion of the eccentricities that survey the allowed parameter 
space eventually reaching $e\to1$
(see panel d in Figure \ref{fig:section}).
Evidently, $e_{\rm max}$ can only increase with  
$N_{ \mbox{\tiny LK}}^{\rm max}$, but we would 
expect that regular or quasi-periodic orbits saturate this growth
after a number of cycles.
We do not see such saturation. 
 
 This figure shows that $N_{ \mbox{\tiny LK}}^{\rm max}$ critically 
 determines the fraction of systems that can merge. Therefore,  
 the binary evaporation or collision times are expected to play a major
 role at setting the merger rates in real astrophysical systems.
 On the other hand, 
 we observe that even after $5-25$ cycles (blue and yellow dashed lines) 
 the effect of the cluster becomes  significant compared to the no cluster
 case (red line): after 25 cycles the number of systems reaching 
  $1-e_{\rm max}\sim10^{-4}$ is boosted by a factor of $\sim20$.
 
 \begin{figure}
\center
\includegraphics[width=8.7cm]{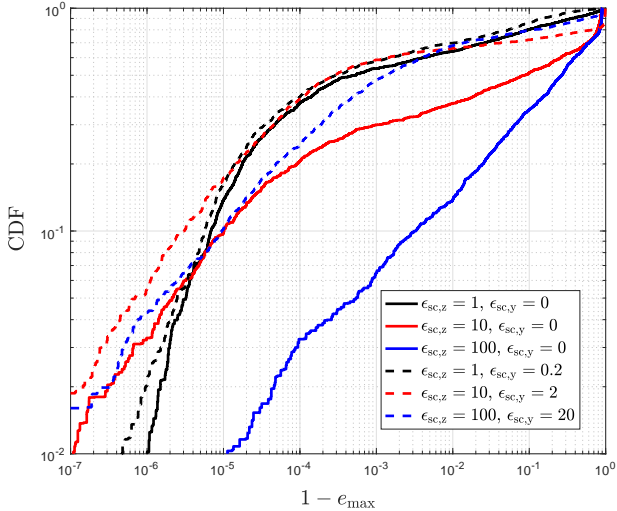}
\caption{Cumulative distribution of the maximum eccentricities 
reached after 100 LK cycles
 for 10,000 randomly oriented binaries  embedded
in a cluster with different values of $\epsilon_{\rm sc,z}$
and $\epsilon_{\rm sc,y}$.
The solid lines show the results for the axysimetric case
($\epsilon_{\rm sc,y}$=0, same as in Figure \ref{fig:population_epsz}).
The dashed lines show the triaxial integrations with
 $\epsilon_{\rm sc,z}/ \epsilon_{\rm sc,y}= \epsilon_{\rm z}/ \epsilon_{\rm y}=5$
and the colors match the runs with the same $ \epsilon_{\rm sc,z}$
for comparison.
We observe that the effect from triaxiality affects the large 
 $\epsilon_{\rm sc,z}$ binaries most dramatically. }
\label{fig:eps_zy}
\end{figure}

\subsection{Fully triaxial case: $0<\epsilon_y\lesssim\epsilon_z$}

We briefly explore the effect from having a triaxial potential
by having a finite value of $\epsilon_{\rm y}$.

In Figure \ref{fig:eps_zy} we show the distribution of maximum eccentricities 
 for a set of integrations with $ \epsilon_{\rm sc,z}=\{1,10,100\}$
in the axisymmetric potential (solid lines with $ \epsilon_{\rm sc,y}=0$)
and triaxial potentials with 
$ \epsilon_{\rm sc,z}/ \epsilon_{\rm sc,y}=\epsilon_{\rm z}/\epsilon_{\rm y}=5$
(dashed lines).
We observe that adding the triaxial component to the axisymmetric
potential increases the fraction of systems reaching higher $e_{\rm max}$
for all $ \epsilon_{\rm sc,z}$
(all dashed lines are above of their corresponding
color-matching solid lines).
This behavior might be expected since adding precession around another
axis can increase the phase-space
volume of the chaotic regions for $\epsilon_{\rm sc,z}=1$ 
and drive the angular momentum $\ell_z={\bf j}\cdot\hat{\bf n}_z$ for the
large $\epsilon_{\rm sc,z}$ cases (see panel c in Figure \ref{fig:eps_y}).

From this figure we also observe that the effect from the triaxial
component at exciting large eccentricities is 
much more dramatic for larger $\epsilon_{\rm sc,z}$.
In particular, for $\epsilon_{\rm sc,z}=1$ the triaxial integrations produce
a distribution of $e_{\rm max}$ that is only slightly above 
its corresponding axisymmetric case.
On the contrary, for  $\epsilon_{\rm sc,z}=100$  we observe 
that the fraction of systems reaching $1-e_{\rm max}\lesssim10^{-4}$
($1-e_{\rm max}\lesssim10^{-5}$) is 
$\simeq25\%$ ($\simeq10\%$) for the triaxial case
compared to only $\simeq3\%$ ($\simeq1\%$) in the 
axisymmetric case. 
This dramatic effect of the triaxial component for the large 
$\epsilon_{\rm sc,z}$ was discussed
in \S\ref{sec:epsy} and it is due to the extra slower precession
around $\hat{\bf n}_y$, which can force
$\ell_z={\bf j}\cdot\hat{\bf n}_z$ to cross $0$ time at which
the eccentricity $e\to1$
(see Figure \ref{fig:eps_y}).

\begin{figure}
\center
\includegraphics[width=8.7cm]{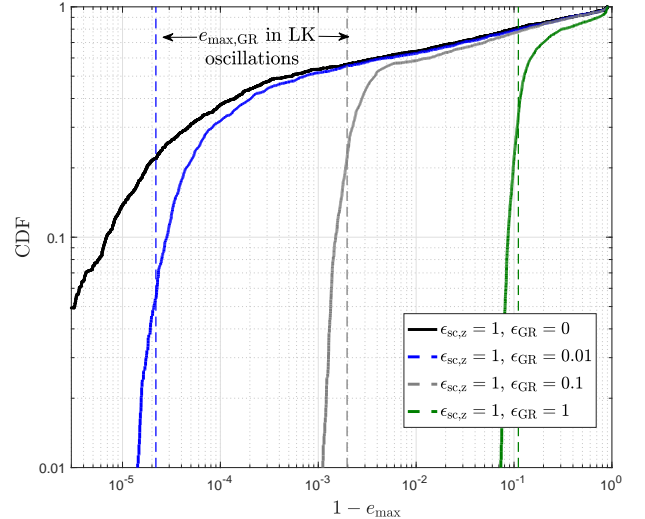}
\caption{Cumulative distribution of the maximum eccentricities 
reached after 100 LK cycles
 for 10,000 randomly oriented binaries  embedded
in a cluster with $\epsilon_{\rm sc,z}=1$ and
different values of $\epsilon_{\rm GR}$ in Equation
(\ref{eq:eps_GR}) as labeled. 
The vertical dashed lines correspond
to the maximum eccentricity reached in LK cycles
in the absence of the cluster perturbations
($e_{\rm max, GR}$ in Equation [\ref{eq:emax_GR}])}
\label{fig:population_GR}
\end{figure}

\subsection{Effect from general relativistic precession}
\label{sec:eps_GR}

It is well known that external sources of apsidal precession such as
 general relativistic precession
can limit the maximum eccentricity growth of the LK
oscillations (e.g., \citealt{blaes02,liu15}).
We quantify the effect from GR precession using the
dimensionless ratio of timescales 
$\epsilon_{\rm GR}=\tau_{ \mbox{\tiny LK}}/\tau_{\rm GR}$  in Equation
(\ref{eq:eps_GR}).

In the absence of the cluster perturbations we can determine the maximum eccentricity 
reached in a LK cycle starting from low eccentricity and $i_{\rm in,out}=90^\circ$
as (e.g., Eq. 35 of \citealt{FT07}):
\ba
e_{\rm max, GR}=\left\{
1-\frac{1}{4}\left[
\left(1+\frac{8}{3}\epsilon_{\rm GR}\right)^{1/2}-1
\right]^2
\right\}^{1/2}.
\label{eq:emax_GR}
\ea

We incorporate the first post-Newtonian corrections in our model
 by adding the following term to the Equation (\ref{eq:edot}):
\ba 
 \frac{d {\bf e}}{d\tau}\bigg|_{\rm GR}=\frac{\epsilon_{\rm GR}}{(1-e^2)^{3/2}}
 \jvec\times\evec.
\label{eq:edot_GR}
\ea

In Figure \ref{fig:population_GR} we show the distribution 
of maximum eccentricities for different values 
of $\epsilon_{\rm GR}$, where $\epsilon_{\rm GR}=0$ corresponds
to our fiducial simulation.
As expected, the larger $\epsilon_{\rm GR}$ (faster GR precession)
the smaller $e_{\rm max}$ from the ensemble of binaries.
For comparison we show the corresponding values of $e_{\rm max, GR}$ from 
Equation (\ref{eq:emax_GR}) for the LK cycles with no cluster.
We observe that the distributions show a clear drop at a maximum 
eccentricity and that this maximum roughly coincides with $e_{\rm max, GR}$.
For instance, if we set $\epsilon_{\rm GR}=0.01$ (0.1) then 
$1-e_{\rm max, GR}\simeq2.2\times10^{-5}$
($\simeq2\times10^{-3}$), while the maximum value in the ensemble of binaries is 
$1-e_{\rm max}\simeq10^{-5}$ ($\simeq0.8\times10^{-3}$).
Thus, from these examples it seems that  $1-e_{\rm max, GR}$ is a 
factor of $\sim2$ larger than the numerical integrations including the 
cluster potential.

In conclusion, relativistic precession quenches the eccentricity 
growth up to a maximum value that is only slightly  higher ($1-e_{\rm max}$ smaller
by a factor of $\sim2$) 
than that derived for the LK mechanism in the absence of the
cluster perturbations ($e_{\rm max, GR}$ in Equation [\ref{eq:emax_GR}]).
In order for BH (stellar) binaries to merge 
we require that $\epsilon_{\rm GR}\lesssim0.01$ 
($\lesssim0.1$) so $1-e_{\rm max}\lesssim10^{-5}$ ($1-e_{\rm max}\lesssim10^{-3}$).

\subsection{Summary of the dynamical results}
\label{sec:summary}

Our main result from the dynamical evolution of binaries
is that when the nodal precession timescale of the binary's center 
of mass due to stellar cluster becomes comparable (within a factor of $\sim10$) 
to the torquing timescale of the binary orbit by to the MBH (Lidov-Kozai timescale), 
the binary often ($\sim10-50\%$ of the cases) reaches extreme eccentricities 
($1-e\lesssim10^{-4}$).
In the absence of the non-spherical cluster potential, this rate decreases by at least 
an order of magnitude.
Thus, we refer to this mechanism as the {\it cluster-enhanced}
Lidov-Kozai mechanism.

We study the dynamical behavior for different values of
$\epsilon_{\rm sc,z}$ (Eq. [\ref{eq:omega_sz}]; ratio between  the LK 
timescales and the nodal precession due to the cluster)
and
identify the following regimes that result in
extreme eccentricity excitation:
\begin{itemize}
\item  for small $\epsilon_{\rm sc,z}$ (slow nodal precession), 
the dynamics resembles the 
octupole-level perturbations in the standard LK mechanism as the main effect
of the cluster is to change the, otherwise conserved, angular momentum 
in the $z-$direction $j_{\rm z}$, which can drive the binary to 
cross $j_{\rm z}=0$ and lead to  $e\to1$ (see Equation
[\ref{eq:jz_evol}]). 
\item for intermediate values of $\epsilon_{\rm sc,z}$,
 the eccentricities and inclinations often undergo widespread chaotic
 evolution as the result of overlapping resonances
 and the eccentricities can diffuse towards $e\to1$ (see surfaces of
 section in Figure \ref{fig:section});
\item for large $\epsilon_{\rm sc,z}$, the binary orbit around the 
MBH undergoes fast nodal precession around $\hat{\bf n}_{\rm z}$ 
creating an effective axisymmetric potential around this axis.
This potential coincides with that from the triple interactions
(LK potential), but defining the orbital elements relative to
$\hat{\bf n}_{\rm z}$ instead of $\hat{\bf j}_{\rm out}$
and changing its amplitude (i.e, the timescale) by a constant factor 
$1-3/2\sin^2 i_{\rm out,z}$.
Thus, the binary undergoes LK cycles with a maximum eccentricity
given by the angle between ${\bf j}$
and $\hat{\bf n}_{\rm z}$ (Eq. [\ref{eq:emax_z}]).
This channel is expected to contribute most significantly to the
extreme eccentricity excitation for
fully triaxial clusters for which 
$\ell_{\rm z}={\bf j}\cdot\hat{\bf n}_{\rm z}$ 
can cross  zero due to an extra slower precession source around
$\hat{\bf n}_{\rm y}$ (see Figure \ref{fig:eps_y}).
\end{itemize}

Other findings include:
\begin{itemize}
\item the dynamics for an axisymmetric cluster
with small $\epsilon_z$ can be well-described by a simple 
two-degree-of-freedom 
Hamiltonian (Equation [\ref{eq:H_BHs}]).
The surfaces of section of this Hamiltonian
show that the LK resonance
overlaps with other resonances arising from
the cluster-driven nodal precession  as 
$\epsilon_{\rm sc,z}$ approaches near unity values
(see Figure \ref{fig:section}).
\item the distribution of maximum eccentricities
depends critically on the maximum evolution
timescale, possibly linked to a chaotic diffusion 
of the eccentricities
(see Figure \ref{fig:population_Nmax});
\item the eccentricity growth is quenched by short-range forces
similar to the LK mechanism. 
In particular, when GR precession is included
the minimum periapsis  distance $a_{\rm in}(1-e_{\rm max})$
in our model is typically a factor of $\sim2$ below 
that from the standard LK mechanism 
($1-e_{\rm max, GR}$ from Equation [\ref{eq:emax_GR}]).
\end{itemize}

We note that, although different in nature, the behavior described
above resembles the dynamics
of four-body systems, where
a fourth body (the nuclear cluster in our case) can change $j_z$
in a triple system \citep{PAST13,hamers15}. 
Recently, \citet{HL17} have studied in more 
detail the dynamics of four-body systems and have
independently arrived to similar conclusions than ours
regarding the general dynamical behavior of triple
systems affected by the nodal precession of a tertiary.

Our dynamical results are expected to have 
important effects on the mergers rates of 
stellar binaries in general.
In what follows, we quantify the relevance our
our mechanism for the compact-object (neutron stars and black holes) 
binaries and differ the study of other combinations of binary
systems for a future study.


\section{Merger of compact-object binaries and 
implications for gravitational wave detections}
\label{sec:BHrate}

In this section we compute the number of compact-object  binary mergers that are formed during 
a star formation episode similar to the one that gave birth to the  young massive stars
observed in the Galactic center. We consider
BH-BH, NS-NS,  and BH-NS binaries.
We use the stellar population synthesis code BinaryStarEvolution \citep[BSE;][]{2002MNRAS.329..897H}  to evolve a population 
of massive binaries stars and compute the properties of compact-object binaries
formed from this population. 
The compact-object binaries are then evolved in a model of the Galactic center
and the dependence of the fraction of merging  binaries on the degree of flattening  of the NC is quantified.
The results of the simulations  are then used as a set of baselines for making predictions about the 
binary merger rates for ground-based laser 
interferometers such as Advanced LIGO and VIRGO.

We assume in what follows that the binary population in galactic nuclei follows the population in the field.
Although little is known about the binary properties in the Galactic center there are observations 
suggesting that our assumption is reasonable.
 Long-term spectroscopic and photometric surveys of the most luminous and massive stars in the vicinity of the massive black hole 
 reveal the existence of  three  OB/WR binaries in the inner 0.2 pc of the Galactic center. 
Using radial velocity change upper limits, \citet{2014ApJ...782..101P}  find that 
the spectroscopic  binary fraction in the Galactic center is $\sim 0.3$,  
 close to that observed in nearby young star clusters \citep{2012Sci...337..444S}. 
These authors also find that the fraction of eclipsing binaries in the central parsec, $\sim 3\%$, 
is  consistent with the fraction of such binaries in local OB star clusters \citep{2009A&A...507.1141L}.
 Overall, observations suggest that the Galactic center binary fraction seems to be similar to the binary fraction in nearby 
 young clusters.

In all our models, we  assumed Solar metallicity and sampled the mass of the most massive star in the  binary
from the initial mass function (IMF) $dN/dm\propto m^{-\alpha}$,
with $\alpha=1.7$ as observed in the galactic center
\citep{2013ApJ...764..155L}, or with the canonical value for field stars of $\alpha=2.3$
\citep{2002Sci...295...82K}.
We adopted a mass ratio $m_2/m_1$ distribution that is uniform between 0 and 1.
 This choice is consistent with the observed mass ratios of massive binary stars 
 \citep{2013ARA&A..51..269D,2012Sci...337..444S,2014ApJS..213...34K}.  Stellar
masses were sampled in the range $10$ to $100\ M_\odot$.
 The upper  limit  on  the  mass  comes  from  the  fact  that  the
stellar evolution tracks used in BSE are not valid above
$100\ M_\odot$.

The distribution of orbital periods  was 
$dN\propto (\log P_{\rm in}/\rm days)^{-0.55}$ in the range $(0.15-4)$, while
the orbital eccentricities were sampled from a 
thermal distribution, $dN\propto e_{\rm in}$. These distributions are both 
consistent with observations 
of  Galactic massive stars in nearby young 
clusters \citep{1991A&A...248..485D,2012Sci...337..444S}.

We evolved the stellar binaries until two compact-objects were formed.
When a compact-object was formed we assigned a natal kick
 from a Maxwellian distribution with dispersion $v_{\rm K}=250\rm km\ s^{-1}$,
in agreement with observational constraints \citep{2005MNRAS.360..974H}.
If the remnant was a BH, the kick velocity was normalized to the mass of the remnant, 
assuming linear momentum conservation. Stars with zero-age main-sequence mass above $40M_\odot$  do not receive any natal kick as in \citet{2001ApJ...554..548F}.

The distribution of the binary orbits around the MBH we adopted is also motivated by observations. 
Spectroscopic studies of the Galactic center reveal a population of $\approx200$ very massive 
  early-type stars, including Wolf-Rayet  stars and O and B type main-sequence stars, giants, and 
supergiants \citep[e.g.,][]{2010ApJ...708..834B,2014ApJ...783..131Y}.
  These massive blue giant stars 
will probably end their lives as an equal number of BHs and NSs. 
  The majority of these massive stars are found at distances of $0.04$ to $0.4\rm pc$ from the MBH,  and  follow
  the surface density distribution $\Sigma \propto R^{-1.5}$ \citep{2010ApJ...708..834B}. About  
  $\approx 20\%$ of them reside in a stellar disk  exhibiting a  clockwise
motion  pattern  on  the  sky \citep{2014ApJ...783..131Y}. 
We assume  that the distribution of the compact-object 
binary orbits follows the distribution
of their stellar progenitors. This is likely a good 
approximation for  systems that 
remain bound after their components 
have received birth kicks \citep{2017arXiv170405850B}.
Accordingly, we sample the semi-major axis of the binary external orbits in 
the range $0.04\leq a_{\rm out}\leq0.4 \ \rm pc$ and 
from the space density distribution $\rho(r)\propto r^{-2.5}$.

To simplify the calculation,  we  suppose  that  the  orbits  of  the  young  stars  randomize in orientation 
and eccentricity on a timescale shorter than the relevant evolutionary timescales.  Several  mechanisms 
might  achieve  this:   e.g., resonant-relaxation, torques from another disk.
Moreover, as also mentioned above, the majority of the young stars do not  
reside inside the disk, but 
are more isotropically distributed around the center
\citep{2014ApJ...783..131Y}.
Accordingly, we assume an isotropic distribution of inclinations, and 
a thermal distribution of 
eccentricities, i.e., $dN\propto e_{\rm out}$. 
Our results are somewhat insensitive to this latter choice
and we checked that by setting $e_{\rm out}=0$ we get similar 
results for the merger fraction.
The longitude of the ascending node, and the argument of periapsis of 
both inner and outer orbit were also randomly
distributed between $0$ and $2\pi$.

 We set $M_{\rm BH}=4\times 10^6\ M_\odot$, and assume that the
field stars follow the density distribution
in Equation (\ref{eq:rho_gc}),
consistent with recent observational constraints \citep{2016ApJ...821...44F,2017arXiv170103817S}.
  The initial conditions were evolved until a maximum timescale $\tau_{\rm max}$,
 which we set equal to  the evaporation timescale $\tau_{\rm evap}$ in Equation
 (\ref{eq:tau_evap}). 
 As discussed in \S\ref{sec:evap}, the binary might significantly 
 change its orbital elements before
 evaporating, so we also provide the results for the much more
 conservative case of setting $\tau_{\rm max}$ equal to the collision 
 time $\tau_{\rm coll}$ in Equation
 (\ref{eq:tau_coll}). 
We stop the integration if the binary periapsis distance to the MBH satisfied  
$a_{\rm out}(1-e_{\rm out})<r_{\rm bt}$,
 with $r_{\rm bt}=a_{\rm in}\left[M_{\rm BH}/(m_1+m_2)\right]^{1/3}$ the binary tidal 
break-up radius.
During the integration, we checked whether $\tau_{\rm LK}\sqrt{1-e_{\rm in}^2}\leq \tau_{\rm GW}$
(see Eq. [\ref{eq:e_merge}]). 
If this condition is satisfied the gravitational merger will
take place within one LK cycle and
the binary was considered to have merged.
We included the effect from relativistic precession using 
Equation (\ref{eq:edot_GR}).
We considered four  models characterized by different values of 
$\epsilon_z=(0.001,0.01, 0.1, 1)$.  
In all models we set $\epsilon_y=0$.

In total we evolved $10^6$ massive binaries. The fraction, $f_2$, of 
these binaries that produced stable BH-BH, NS-NS, and BH-NS binaries are  given in Table~\ref{TF}. 
Only a small fraction of massive 
binary stars produce a 
bound pair of compact objects; most binaries are either disrupted due to  natal kicks
or merge during their main-sequence evolution.

Table\ \ref{TF} summarizes the results of our calculations, showing the fraction of formed BH binaries that merge, $f_{\rm m}$,  as a function of the parameter $\epsilon_z$ that measures
flattening of the cluster potential (see Equation [\ref{eq:ellip}] for its relation to cluster
ellipticity $\mathcal{E}$ for $r\sim r_0$). 
When computing $f_{\rm m}$ we did not include binaries that underwent a phase of common-envelope evolution,
since these binaries have $\epsilon_{\rm GR}>1$ and the LK process is strongly  quenched. Our rate estimates below are therefore 
only for mergers due to the cluster-enhanced LK mechanism, and do not include those binaries that would merge as a result o stellar evolution processes.

The  results  shown in Table\ \ref{TF}  exemplify the 
importance of the mechanism described in this paper, by showing a strong dependence of the merging fraction on the  NC morphology. 
For BH-BH binaries we show that
nearly spherical models ($\epsilon_z=0.001$ or ellipticity 
$\mathcal{E}\simeq0.001$) the merger
 fraction is only $\approx 10^{-3}$.
We expect therefore  a small merger rate of BH binaries in spherical NCs, in agreement with previous work \citep{2012ApJ...757...27A}.
In contrast, for $\epsilon_z= 0.1$ ($\mathcal{E}\simeq0.1$) 
the merging fraction increases up to $\approx 6\%$, and for $\epsilon_z= 1$ (fully flattened,
$\mathcal{E}=1$) to $\approx 17\%$. 
For BH-NS and NS-NS binary mergers we observe that no mergers occur for the spherical 
clusters, while the fraction reach $f_{\rm m}\sim1\%$ for flattened clusters.

We remark that we start evolving the orbits of the compact-object binaries 
under the gravitational influence from the MBH and the cluster once they form.
One potential concern from this simplification is that binaries might
merge during the main sequence before they become compact objects 
(e.g., \citealt{ATH17}). We have checked, however, 
that most BH-BH mergers, $ 77\%$ and $75\%$ for $\epsilon_z=0.1$ and 1 respectively, occurred after
a time that was longer than the main sequence lifetime of both binary components, 
meaning that these might have avoided an unwanted 
merger prior to the formation of the two BHs. Having relatively late mergers
(after many LK timescales and continuous in time) is expected from the chaotic 
diffusion of the eccentricities (see Figure \ref{fig:population_Nmax}).

We conclude that allowing for a more realistic model in which the shape of the NC
is not perfectly spherical will significantly increase the merger rates of compact-object binaries, 
which for BH-BH binaries it does so by factor of $\sim 30-100$ for $\epsilon_z\gtrsim0.1$
(ellipticities $\gtrsim0.1$).

\begin{table}
  \caption{Results of the population synthesis models of massive binary stars.}
  \centering
\begin{tabular}{llllllll}
\hline  
\hline
 & $\epsilon_z$ &  $f_2(\%)$  &   $f_2(\%)$ & $f_{\rm m}(\%)$ & $f_{\rm m}(\%)$ \\
& &$\alpha=1.7$  & $\alpha=2.3$ & $t<\tau_{\rm coll}$ & $t<\tau_{\rm evap}$    \\ 
  \hline 
BH-BH & 0.001 & 4.5 & 2.5 & 0.17 & 0.17 \\
... & 0.1 & 4.5 & 2.5  & 1.5 & 5.8 \\
... & 0.3 & 4.5 & 2.5  & 2.6 & 12 \\
... & 1 & 4.5 & 2.5    & 3.9 & 17 \\
\hline
BH-NS & 0.001 & 1.3 & 0.87 & 0 & 0   \\
... & 0.1 & 1.3 & 0.87 &  0.072 &  0.51        \\
... & 0.3 & 1.3 & 0.87  &  0.26 &  1.2 \\
... & 1 & 1.3 & 0.87    &  0.29 &  1.5 \\
\hline
NS-NS & 0.001 & 0.22 & 0.28  &  0    &  0 \\
... & 0.1 & 0.22 & 0.28   & 0.045     & 0.045           \\
... & 0.3 & 0.22   & 0.28 & 0.18 & 0.27         \\
... & 1 & 0.22     & 0.28 & 0.41 & 0.50 \\
  \hline \hline
\end{tabular}
\\ {Fraction of massive binaries that produce a compact-object binary at the end of their evolution ($f_2$), and 
the fraction of the formed binaries that merged due to the cluster-enhanced LK mechanism
($f_{\rm m}$)
before a time $t$.
These quantities are given for different values of flattening of the cluster potential
$\epsilon_z$ and type of compact-object binaries.
\\ 
  }\label{TF}
\end{table}

\subsection{Merger rate estimates}

The total mass of the young stars currently observed 
in the inner $\approx 1\rm pc$ region is a few times $10^4\ M_{\odot}$ \citep{2013ApJ...764..155L}.
The starburst that created this
population of massive stars may have been just the most recent episode of an
ongoing process of continuous star formation.   Observations 
suggest   in fact that  a large fraction
of the Milky Way NC was built up gradually in time through in-situ star formation episodes similar to the one from which the observed massive stars were formed \citep{2004ApJ...601..319F,2011ApJ...741..108P}. 
Young stellar populations are ubiquitous in NCs that are 
found in external galaxies \citep[e.g.,][]{2015AJ....149..170C}, suggesting 
continuous/episodic star formation as a generic mode for NC formation 
and growth \citep{2006AJ....132.1074R,2015ApJ...812...72A,2015ApJ...806L...8A}.
 
 Observations of the central regions of the Milky Way show
that this region is dominated by a dense NC consisting of a population of old
stars with a total mass of $\sim 3\times10^7M_{\odot}$
\citep{2014A&A...566A..47S,2017arXiv170103817S,2017MNRAS.466.4040F}
with most young stars residing  inside $\sim0.5$ pc 
\citep{2014A&A...566A..47S,2017arXiv170103817S,2017MNRAS.466.4040F}.
We  assume here that a fraction of this mass was formed 
via continuous in-situ star formation throughout
the age of the Galaxy. Under this assumption we compute the rate at which compact objects
must form, $\Gamma_{\rm co}$, such that the total mass formed  in stars and stellar remnants
after 10 Gyr is $10^7M_\odot$, approximately the total stellar mass within the sphere 
of influence of SgrA*.
The value of $\Gamma_{\rm co}$ depends on the assumed IMF and on the star formation history of the NC.
Table 1 in \citet{2010MNRAS.402..519L} gives  the composition of an evolved stellar population
for different star formation histories and IMFs; we use this table to compute $\Gamma_{\rm co}$.

For a top-heavy IMF similar to the one adopted here and a constant star formation,
\citet{2010MNRAS.402..519L} find 
that $10^5$ BHs and NSs are formed for every 
$10^6 M_\odot$ in stars and stellar remnants. 
If we require that $10^7 M_\odot$ in stars and stellar remnants are formed after 10 Gyr,
this gives $\Gamma_{\rm co}\sim 10^6/10\,\rm{Gyr}=10^{-4}\rm yr^{-1}$.
In other words, the predicted total number
of compact objects formed in the NC after $10$ Gyr in this case is $\sim 10^6$.
For a canonical IMF and a constant star formation,
\citet{2010MNRAS.402..519L} find that  
$1.7 \times 10^4$ BHs and NSs are formed for every 
$10^6 M_\odot$ in stars and stellar remnants, which gives 
$\Gamma_{\rm co}\sim 2\times 10^{-5}\rm yr^{-1}$.
Thus, in this latter case the predicted total number
of compact objects formed in the NC after 10 Gyr is $\sim 2\times 10^5$.
The number of compact objects predicted to form by a top-heavy IMF is 
therefore approximately 
10 times larger than for a canonical IMF, given an equal total mass 
formed after 10 Gyr.

Given the rate of formation of compact objects in the central cluster, 
we can compute the merger rate of compact-object binaries  as:
\begin{eqnarray}\label{rate}
  \Gamma_{\rm m}= \Gamma_{\rm co} f_{\rm 2} f_{\rm m}\ ,
\end{eqnarray}
where we have assumed that all massive stars form in binaries, similarly 
to what found from observation of nearby young clusters \citep{2012Sci...337..444S}.

Next, we estimate the 
merger rate for compact-object binaries.
In order to do so we further assume that all NCs are characterized by some degree of spherical asymmetry 
(ellipticities $\gtrsim0.1$ or axis ratios $q\lesssim0.9$). This assumption appears
 to be reasonable. 
 The best-fitting Schwarzschild model of the Galactic center in 
 \citet{2017MNRAS.466.4040F} has  an axis ratio $q\sim0.3$ in the nuclear region, increasing up to $\sim0.8$ at $\sim1-2$ pc.
Similarly, \citet{2016ApJ...821...44F} measure a mean axis ratio of $q\sim0.8$.
\citet{2014A&A...566A..47S} find $q\sim0.7$ from Spitzer/IRAC photometry.
NCs in external galaxies are also often observed to be strongly flattened, despite of
projection effects that can make the clusters look more spherical. 
The nuclei  of the nearby galaxies M31, M32, and M33 all have substantial ellipticities $\gtrsim 0.2$ \citep{1998AJ....116.2263L}, while 
the majority of the NCs in the galaxy samples of  \citet{2014MNRAS.441.3570G} and of \citet{2015AJ....149..170C}
have measured ellipticities.
Typical values are of order $\approx 0.2-0.3$ but values up to $\approx 0.9$ are also found. 

Finally, we stress that the analysis presented here
makes several simplifying  assumptions
which will be relaxed and detailed  in future work.
 Here, we simply note that the  merger rates we derived should be
 viewed as an order-of-magnitude estimate
indicating that the cluster-enhanced LK mechanism
is an important new merger channel for 
compact-object binaries, possibly competing with 
 traditional scenarios which
 invoke either isolated binary evolution, 
or dynamical formation in globular clusters.

\begin{figure}[t!]
\center
\includegraphics[width=8.7cm]{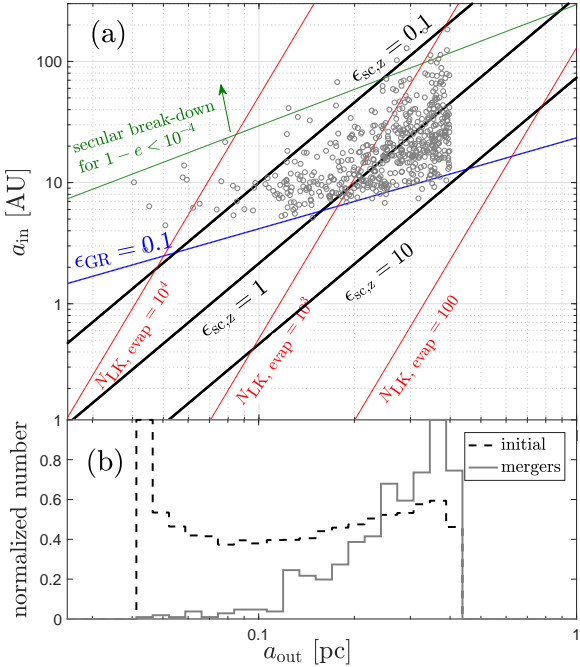}
\caption{
{\it Panel a}: 
semi-major axes $a_{\rm in}$ and $a_{\rm out}$ of the black hole 
binaries that merge in our population synthesis study (empty circles).
The cluster potential in Equation (\ref{eq:phi_r}) is assumed to be axisymmetric 
with $\epsilon_{\rm z}=0.1$ and the spherical component of the density 
profile is given by the Equation (\ref{eq:rho_gc}).
The various lines are the same as in Figure 2
and are computed using $m_1+m_2=20M_\odot$
(roughly the mean and median of the merged systems):
$\epsilon_{\rm sc,z}=\tau_{\mbox{\tiny LK}}/\tau_{\rm sc,z}$
(Eq.  [\ref{eq:omega_sz}], black lines),
$N_{ \mbox{\tiny LK, evap}}$ is the number of LK cycles allowed before 
the binary evaporates (Eq. [\ref{eq:Nevap}], red lines), 
and $\epsilon_{\rm GR}=\tau_{\mbox{\tiny LK}}/\tau_{\rm GR}=0.1$ 
is an approximate boundary above which GR quenches extreme eccentricity
excitation (Eq. [\ref{eq:eps_GR}], blue line),
and the green line corresponds to the limit above which 
the secular approximation breaks down when 
$1-e\lesssim10^{-4}$ from Equation (18) in
\citet{AMM2014}.
{\it Panel b}: normalized histogram of $a_{\rm out}$
for the initial distribution (dashed black lines) and
for the systems that merge (solid gray lines).}
\label{fig:mergers}
\end{figure}

\subsection{BH-BH binaries}
BH-BH are the most common form of compact-object binaries produced in our models.

In panel a of Figure \ref{fig:mergers}, we show the 
semi-major axes $a_{\rm in}$ and $a_{\rm out}$
of the BH-BH binaries that merge in our simulations
and also show the relevant timescale ratios from panel b of Figure 2.
We note that mergers happen preferentially for 
$\epsilon_{\rm sc,z}\sim1$, which skews the distribution 
of $a_{\rm out }$ to wider separations: compare initial distribution
(black dashed line) on panel b with those that merge (gray line).
We also observe that the merged systems are expected to undergo 
changes in their orbital parameters in secular timescales only (green line
in panel a) and the dynamics is, therefore, properly captures by our 
equations of motion.

Taking the merging fraction, $f_{\rm m}$, from Table\ \ref{TF}, 
we obtain from Equation\ (\ref{rate}) a merger rate 
in the range $\Gamma_{\rm m} \approx (0.03,\ 0.8)\rm\ Myr^{-1}$,
where the lower limit corresponds to 
$\epsilon_z=0.1$, and a canonical IMF;
the upper limit is for $\epsilon_z=1$, and  a top-heavy IMF.
This merger rate can be compared to that corresponding 
to standard binary formation 
models. These latter models predict that the 
evolution of isolated binary stars in the Galactic field results in a 
merger rate  in the range $0-30\rm\ Myr^{-1}$, assuming
Solar metallicity \citep{2012ApJ...759...52D}.
We conclude that our new mechanism gives a local merger rate for BH-BH binaries 
that is comparable or even larger to that predicted by other
standard scenarios \cite[e.g.,][]{2016ApJ...818L..22A}.

In order to estimate a rate per volume we use
a number density of  galaxies of $\approx 0.02\rm\ Mpc^{-3}$ \citep[e.g.,][]{2005ApJ...620..564C,2008ApJ...675.1459K}.
This gives a merger rate in the range
\begin{equation}
\Gamma_{\rm BH-BH} \approx (0.6,\ 15)\rm\ Gpc^{-3}yr^{-1}\ .
\end{equation}
Realistic values are likely to be somewhere in the middle of this range.
For instance,  by taking  $\epsilon_z=0.1$
we have  $\Gamma_{\rm BH-BH} \approx 5\rm\ Gpc^{-3}yr^{-1}$ for a top-heavy initial mass function.
If we take  $\epsilon_z=0.3$, we have $\Gamma_{\rm BH-BH} \approx 1\rm\ Gpc^{-3}yr^{-1}$ for a canonical initial mass function.

Remarkably, the merger rate we derived 
is similar to the predicted local merger rate of BHs 
formed in globular clusters $\Gamma = (2,\ 20) \rm\ Gpc^{-3}yr^{-1}$ \citep{2016PhRvD..93h4029R}.
This result can appear surprising: according to our models, the 
cluster-enhanced LK mechanism  will realistically 
 only produce  about one thousand  BH mergers over 10 Gyr. 
Over the same timescale one globular cluster alone will produce
$\approx 100$ BH mergers, resulting in a total of $\approx 2\times 10^4$  mergers
produced by  $200$ globular clusters in the Galaxy (C. Rodriguez; private communication).  
Part of the reason why the globular cluster rate is 
relatively low 
is because a large fraction of the mergers happen early on in the evolution at high redshift, i.e., the merger
rate decreases considerably  with time \citep[see Figure 12 in][]{2016PhRvD..93h4029R}.
Our more realistic assumptions result in $\lesssim 1/10\rm th$ of the BH mergers produced in globular clusters,
however, these are produced at a $constant$ rate giving a relatively higher 
local merger rate.

Finally, we note that  \citet{2016ApJ...831..187A} derived a merger rate 
in nuclear star clusters without central massive black holes of
$\sim 1\rm\ Gpc^{-3}yr^{-1}$, where the BH binaries form 
dynamically as in GCs.
Our estimates show that even when a MBH is present in the NC,
BH-BH binaries can still form and merge at a high rate.

\subsection{BH-NS mergers}
About $1\%$ of the massive binaries we evolved produced a BH-NS binary. 
Taking the merging fraction, $f_{\rm m}$, from Table\ \ref{TF}, 
we obtain from Equation\ (\ref{rate}) a Galactic  merger rate  
in the range $\Gamma_{\rm m} \approx (10^{-3},\ 2\times 10^{-2})\rm\ Myr^{-1}$,
where, as before, the lower limit corresponds to 
$\epsilon_z=0.1$, and a canonical IMF, while
the upper limit is for $\epsilon_z=1$, and  a top-heavy IMF.
For comparison, the standard binary formation channel  produces a 
Galactic merger rate of BH-NS binaries in the range $0-10\rm\ Myr^{-1}$, assuming
Solar metallicity \citep{2012ApJ...759...52D}.  Dynamical formation in globular  
clusters is unlikely to result in an important number of BH-NS binaries, as NSs do not efficiently segregate 
to the center, and most of them are ejected at birth
from the parent clusters due to their high natal kicks.

The corresponding merger rate for BH-NS binaries per unit volume due to the cluster-enhanced LK mechanism 
is in the range
\begin{equation}
\Gamma_{\rm BH-NS} \approx (2\times 10^{-2},\ 0.4)\rm\ Gpc^{-3}yr^{-1}\ .
\end{equation}
 
\subsection{NS-NS mergers}
Only about $0.3\%$ of massive binaries in our models form 
a NS-NS binary. 
The merging fractions given in Table\ \ref{TF}, 
lead to a Galactic  merger rate  
in the range $\Gamma_{\rm m} \approx (3\times 10^{-5},\ 10^{-3})\rm\ Myr^{-1}$,
where, as before, the lower limit corresponds to 
$\epsilon_z=0.1$, and a canonical IMF, and
the upper limit is for $\epsilon_z=1$, and a top-heavy IMF.
The resulting merger rate per unit volume for NS-NS binaries 
predicted by our models is $\Gamma_{\rm NS-NS}\lesssim 0.02 \rm\ Gpc^{-3}yr^{-1}$,
and it is therefore negligible compared with the predicted merger rates  from
the evolution of isolated binaries in the Galactic field \citep[e.g.,][]{2012ApJ...759...52D}. 

We remark that in our calculation we have not included mergers favored
by stellar evolution processes that can occur prior to the formation of the two NSs.
Evolution driven by mass transfer 
\citep[e.g.,][]{2016ApJ...825...71D,2016ApJ...825...70D},
followed by a common envelope phase \citep[e.g.,][]{1998ApJ...493..351K}  
can lead to the formation of short period 
NS-NS binaries which can merge in less than one Hubble time \citep[e.g.,][]{2001ApJ...556..340K,2015ApJ...814...58D}.
For NS-NS binaries, we expect these latter type of mergers
 to be more frequent than those due to the cluster-enhanced LK mechanism.

\section{Conclusions}
\label{sec:conclusion}

Galactic centers are often
observed to host a central massive
black hole embedded by a non-spherical nuclear
star cluster.
These clusters are composed by a wide range of stellar populations, 
including very massive stars.

We have studied the orbital evolution of stellar binaries 
in these environments, focusing on the excitation
of large eccentricities that can lead to mergers
of compact-object binaries by gravitational 
radiation. 

 We show that the excitation of extremely large  eccentricities
 depends critically on the ratio between the Lidov-Kozai (LK)
 timescale due to the MBH  and the timescale of the
 nodal precession of the binary around the MBH due to an axisymmetric
 (or triaxial)  star cluster, 
 which we define as $\epsilon_{\rm sc,z}$ (Eq. [\ref{eq:omega_sz}]).
 
 For $\epsilon_{\rm sc,z}\sim0.1-10$,
 reached by a wide range of binaries in the
 galactic center, we find that $\sim10-50\%$ of randomly-oriented binaries
reach $1-e\sim10^{-4}-10^{-6}$, compared to only
 $\sim0.1-1\%$ in the idealized case of an spherical star cluster
 ($\epsilon_{\rm sc,z}=0$). 
Thus, we identify the {\it cluster-enhanced}
Lidov-Kozai mechanism as a efficient new channel to
drive binary mergers in galactic nuclei.

We study the dynamics for different values 
of $\epsilon_{\rm sc,z}$
and identify two main channels for the
extreme eccentricity excitation:
(i) chaotic diffusion of the eccentricities due to the 
overlap of the LK resonance and resonances arising from the 
cluster-induced nodal precession;
(ii) cluster-driven variations of the mutual inclinations
between the binary orbit
and its center of mass orbit around the MBH, which drive 
the  otherwise conserved LK constant of motion $j_z$
to zero.

We have carried out a population synthesis study
to estimate the merger rate of compact-object 
binaries in NCs hosting a central MBH.
Under the observationally motivated assumption that most nuclear star 
clusters have finite ellipticities ($\gtrsim0.1$),
we derived the following merger rates:
(i) BH-BH binaries in the range $(0.6,\ 15)\rm\ Gpc^{-3}yr^{-1}$;
(ii) BH-NS binaries in the range $(2\times10^{-2},\ 0.4)\rm\ Gpc^{-3}yr^{-1}$. 
These merger rates are comparable to those from 
either isolated binary evolution
or dynamical formation in globular clusters.
The wide range of predicted rates
is explained mainly by the uncertainties
in the star formation history
of the galactic nuclei.
 
Our results suggest a strong correlation between 
 the merger rates of black hole  binaries in
galactic nuclei and the
morphology 
of their central clusters.

\acknowledgements 
We thank the referee Ben Bar-Or for providing with a careful and constructive
report that has helped us to improve the presentation of 
our paper.
We thank Adrian Hamers for sharing his related 
work on quadruples and commenting on  an early version
our paper and Scott Tremaine for providing useful feedback 
on the manuscript.
We are grateful to Chris Matzner, Dan Tamayo,
Diego Mu\~noz,  Fred Rasio, Juna Kollmeier, 
Katerina Chatziioannou, Norm Murray, Renu Malhotra,
Yanqin Wu, and Aaron Zimmerman 
for stimulating discussions and comments.
CP acknowledges support from the Gruber Foundation. FA acknowledges 
support from a CIERA postdoctoral fellowship at Northwestern University.

\appendix
\section{Equations of motion for a binary embedded in a nuclear star cluster with 
a central massive black hole}

We derive the orbit-averaged equations of a motion from the 
potential per unit mass in Equation (\ref{eq:phi_r}):
\ba 
{\Phi}({\bf r})&=&-\frac{GM_{\rm BH}}{r}+\frac{4\pi G}{(3-\gamma)(2-\gamma)}
\rho_{s,0} r_0^2\left(\frac{r}{r_0}\right)^{2-\gamma}
\left[1+
\epsilon_z\frac{(\hat{\bf n}_z\cdot {\bf r})^2}{r^2}+
\epsilon_y\frac{(\hat{\bf n}_y\cdot {\bf r})^2}{r^2}\right].
\ea

We start by averaging over the outer orbit the terms that appear on the triaxial 
potential. 
First, as shown in \citet{ivanov05} and \citet{MV11} the averaging of the 
spherical component results in
\ba
\bigg<\left(\frac{r}{r_0}\right)^{2-\gamma}\bigg>=
\left(\frac{a_{\rm out}}{r_0}\right)^{2-\gamma}
~ _2F_1\left[-\frac{2-\gamma}{2},-\frac{3-\gamma}{2},1,e^2\right]
\simeq\left(\frac{a_{\rm out}}{r_0}\right)^{2-\gamma}
\left(1+\alpha(\gamma)-\alpha'(\gamma)j_{\rm out}^2\right),
\ea
where $_2F_1$ is the ordinary hypergeometric function and
the constants terms can be approximated as (e.g., \citealt{MV11}):
\ba
\alpha(\gamma)\simeq  \frac{3}{2}-\frac{79}{60}\gamma+\frac{7}{20}\gamma^2-
\frac{1}{30}\gamma^3
~~{\rm and }~~
\alpha'(\gamma) \simeq  \frac{3}{2}-\frac{29}{20}\gamma+\frac{11}{20}\gamma^2-
\frac{1}{10}\gamma^3
\ea
for $0\leq\gamma<2$.

Second, the non-spherical in the potential involves the averaging of the following terms:
\ba 
r_{\rm out}^{-\gamma}(\hat{\bf n}_i\cdot {\bf r}_{\rm out})^2=r_{\rm out}^{2-\gamma}\cos^2 \phi 
\left(\hat{\bf n}_i\cdot {\bf \hat{e}_{\rm out}}\right)^2
+2r_{\rm out}^{2-\gamma}\sin \phi\cos \phi \left(\hat{\bf n}_i\cdot {\bf \hat{e}}_{\rm out}\right)
\left(\hat{\bf n}_i\cdot {\bf \hat{q}}_{\rm out}\right)
+r_{\rm out}^{2-\gamma}\sin^2 \phi \left(\hat{\bf n}_i\cdot {\bf \hat{q}}_{\rm out}\right)^2,
\label{eq:rout_phi}
\ea
where $\phi$ is the azimuthal angle in the 
${\bf \hat{e}}_{\rm out}-{\bf \hat{q}}_{\rm out}$
plane indicating the position of the bodies along their 
Keplerian orbits.
In principle, we can average the above expression
over one period of this phase 
angle $\phi$ arbitrary value of $\gamma$ (see e.g., \citealt{ST99,TY14} for the averaging
method), but the mathematical expressions are complicated and not very illuminating.
Instead, we provide the explicit expression for 
 $\gamma=0$ (uniform ellipsoid) and 
$\gamma=1$ in which cases the algebra is greatly simplified
and they capture the dynamical behavior for the general case.
We note that the previous study by \citet{MV11}, based on the previous work
by \citet{ST99} and \citet{SS00} using Delaunay variables, modeled the 
cluster potential as the sum of an spherical cusp for an arbitrary $\gamma$
and a triaxial uniform ellipsoid ($\gamma=0$), which is somewhat more
general at the expense of having to define an extra density profile
for the ellipsoid. Our approach  and that in \citet{MV11} should give
similar results.

For $\gamma=0$ the averaging of Equation (\ref{eq:rout_phi}) results in:
\ba 
\big<(\hat{\bf n}_i\cdot {\bf r}_{\rm out})^2\big>=\frac{a_{\rm out}^2}{2}
\left [j_{\rm out}^2+5\left(\hat{\bf n}_i\cdot {\bf e}_{\rm out}\right)^2
-\left(\hat{\bf n}_i\cdot {\bf j}_{\rm out}\right)^2 \right],
\ea 
implying that the averaged potential due to the cluster becomes
\ba 
\bar{\Phi}_{\rm s} |_{\gamma=0}&=&\frac{1}{3}\pi G\rho_{s,0} a\out^2
\left\{5-3j_{\rm out}^2+\sum_{i=y,z} \epsilon_i 
\left[j_{\rm out}^2+5\left(\hat{\bf n}_i\cdot {\bf e}_{\rm out}\right)^2
-\left(\hat{\bf n}_i\cdot {\bf j}_{\rm out}\right)^2 \right]\right\}.
\label{eq:phi_gamma0}
\ea

For $\gamma=1$ the averaging of Equation (\ref{eq:rout_phi}) results in:
\ba 
\bigg<\frac{(\hat{\bf n}_i\cdot {\bf r}_{\rm out})^2}{r_{\rm out}}\bigg>=\frac{a_{\rm out}}{2}
\left [j_{\rm out}^2+3\left(\hat{\bf n}_i\cdot {\bf e}_{\rm out}\right)^2
-\left(\hat{\bf n}_i\cdot {\bf j}_{\rm out}\right)^2 \right],
\ea 
implying that the averaged potential due to the cluster becomes
\ba 
\bar{\Phi}_{\rm s} |_{\gamma=1}&=&\pi G\rho_{s,0} r_0^2
\left(\frac{a_{\rm out}}{r_0}\right)
\left\{3-j_{\rm out}^2+\sum_{i=y,z}  \epsilon_i 
\left[j_{\rm out}^2+3\left(\hat{\bf n}_i\cdot {\bf e}_{\rm out}\right)^2
-\left(\hat{\bf n}_i\cdot {\bf j}_{\rm out}\right)^2 \right]\right\}.
\label{eq:phi_gamma1}
\ea

In what follows, we present the results for $\gamma=1$ only, while
those for $\gamma=0$ follow trivially.
Following a similar procedure for the other term in the potential
we can then write the averaged potential over the inner 
and outer orbits:
\ba
\bar{\Phi}&\simeq&\frac{3GM_{\rm BH}\mu_{\rm in} a^2}
{4a_{\rm out}^3(1-e_{\rm out}^2)^{3/2}}
\left[  \frac{1}{6}-e_{\rm in}^2
-\frac{1}{2}({\bf {j}}_{\rm in}\cdot \hat{\bf j}_{\rm out})^2+
\frac{5}{2}  ({\bf e}_{\rm in}\cdot \hat{\bf j}_{\rm out})^2\right]\nonumber\\
&&+
\pi  \mu_{\rm out}G\rho_{s,0} r_0^2
\left(\frac{a_{\rm out}}{r_0}\right)
\left\{3-j_{\rm out}^2+\sum_{i=y,z}  \epsilon_i 
\left[j_{\rm out}^2+3\left(\hat{\bf n}_i\cdot {\bf e}_{\rm out}\right)^2
-\left(\hat{\bf n}_i\cdot {\bf j}_{\rm out}\right)^2 \right]\right\}.
\label{eq:phi_full}
\ea
where $\mu_{\rm in}=m_1m_2/(m_1+m_2)$ and 
$\mu_{\rm out}=(m_1+m_2)M_{\rm BH}/(m_1+m_2+M_{\rm BH})$
are the reduced masses of the inner and outer orbits, respectively.
In order to simplify the dynamics we have ignored the effect from the cluster on the inner orbit, which
is a reasonable approximation only well-inside the sphere of influence of the MBH.
In particular, for $\gamma=0$ the ratio between the amplitude of potential due to the cluster
and that from the MBH acting in the inner orbit is $M_s(a\out)/(3M_{\rm BH})\ll1$, where
 $M_s(a\out)$ is the enclosed mass of stars in the cluster in a sphere of radius $a\out$. 
Incorporating the potential of the cluster acting in the inner orbit 
is straightforward and it simply follows
from exchanging the outer orbit by the inner one in the
Equations (\ref{eq:phi_gamma0}) and (\ref{eq:phi_gamma1}).

The first term in the potential corresponds to the well-known Lidov-Kozai interaction 
potential only up to quadrupole level. We have avoided writing the octupole term in the potential
to keep the equations simpler, but they are included 
in our code and
can be found elsewhere in the vectorial formalism
(e.g., \citealt{BF14,liu15,2015ApJ...805...75P,PM17}).
We note, however,  that in the regime that we explore in our paper with
$a_{\rm in}\sim1-10$ AU  and $a_{\rm out}\sim0.1-1$ pc the octupole term 
has a negligible effect on the dynamics because (i) $\epsilon_{\rm oct}$ in Equation
(\ref{eq:oct}) is generally too small ($10^{-5}-10^{-3}$), while mostly values $\gtrsim0.01$
are likely to contribute significantly to the dynamics \citep{li14}; (ii) the spherical cluster 
induces fast apsidal precession so the octupole contribution, which depends
on the apsidal orientation of the outer orbit 
($\propto {\bf e}_{\rm in}\cdot{\bf e}_{\rm out} $),
 averages out to zero (see Eq. [\ref{eq:oct_sc}] for reference).

Finally, the equations of motion can then be conveniently written in a 
dimensionless form as (e.g., \citealt{M39,TTN09}):
\ba
 \frac{d {\bf e}}{d\tau}&=& 2\jvec\times\evec
 -5\left(\evec\cdot\hat{\bf j}_{\rm out}\right)\jvec\times\hat{\bf j}_{\rm out}+
\left(\jvec\cdot\hat{\bf j}_{\rm out}\right)\evec\times\hat{\bf j}_{\rm out}	,  
\label{eq:edot}\\
\frac{d {\bf j}}{d\tau}&=&
 \left({\bf j}\cdot\hat{\bf j}_{\rm out}\right){\bf j} \times \hat{\bf j}_{\rm out}
-5\left({\bf e}\cdot\hat{\bf j}_{\rm out}\right){\bf e}\times\hat{\bf j}_{\rm out},
\label{eq:jdot}\\
 \frac{d {\bf e}_{\rm out}}{d\tau}&=& \epsilon_{\rm sc} \bigg\{
 {\bf e}_{\rm out}\times{\bf j}_{\rm out}
 -\sum_{i=y,z} \epsilon_i\big[
3 \left(\hat{\bf n}_i\cdot {\bf e}_{\rm out}\right)	{\bf j}_{\rm out}  \times \hat{\bf n}_i+
	 {\bf e}_{\rm out} \times {\bf j}_{\rm out}
-\left(\hat{\bf n}_i\cdot {\bf j}_{\rm out}\right){\bf e}_{\rm out} \times\hat{\bf n}_i
	\big]\bigg\},  
	\label{eq:eout_evol}\\
\frac{d {\bf j}_{\rm out}}{d\tau}&=&\epsilon_{\rm sc} \sum_{i=y,z} \epsilon_i\big[
\left(\hat{\bf n}_i\cdot {\bf j}_{\rm out}\right){\bf j}_{\rm out}\times
 \hat{\bf n}_i
 -3 \left(\hat{\bf n}_i\cdot {\bf e}_{\rm out}\right)	 
{\bf e}_{\rm out} \times \hat{\bf n}_i
 \big], 
 	\label{eq:jout_evol}
\ea
where the unit time is given in Lidov-Kozai timescales, $\tau=t/\tau_{\mbox{\tiny LK}}$,
with
\ba
\tau_{ \mbox{\tiny LK}}=  \frac{(m_1+m_2)}{M_{\rm BH}}
\frac{a_{\rm out}^3(1-e_{\rm out}^2)^{3/2}}{a^3}\frac{2P}{3\pi},
\ea
and we have defined $\epsilon_{\rm sc}\equiv\tau_{ \mbox{\tiny LK}}/\tau_{\rm sc}$ with 
\ba 
 \tau_{\rm sc}^{-1}=G P_{\rm out} \rho_{s}\left(\frac{r_0}{a_{\rm out}}\right)\simeq
 G P_{\rm out} \rho_s(a_{\rm out})
 \label{eq:tau_sc}
\ea
where $\rho_s(a_{\rm out})\equiv\rho_s(|{\bf r}|=a_{\rm out})$ with $\epsilon_z=\epsilon_y=0$ or the density for spherical component.
Thus, the ratio of timescales $\epsilon_{\rm sc}$ (sc stands for stellar cluster), 
assuming  $\epsilon_z,\epsilon_y\ll1$
and $m_1+m_2\ll M_{\rm BH}$,  becomes
\ba
\epsilon_{\rm sc}\equiv\frac{\tau_{\mbox{\tiny LK}}}{\tau_{\rm sc}}&=&
\frac{8\pi}{3}\left(\frac{\rho_s(a_{\rm out})a_{\rm out}^3}{M_{\rm BH}}\right)
\left[\frac{a_{\rm out}(1-e_{\rm out}^2)}{a}\right]^{3/2}\left(\frac{m_1+m_2}{M_{\rm BH}}\right)^{1/2}
\label{eq:epsilon_s}\\
&=&\frac{4}{3}\left(\frac{M_s(a_{\rm out})}{M_{\rm BH}}\right)
\left[\frac{a_{\rm out}(1-e_{\rm out}^2)}{a}\right]^{3/2}\left(\frac{m_1+m_2}{M_{\rm BH}}\right)^{1/2}
\label{eq:omega_s}
\ea
where we have defined the enclosed mass with $\epsilon_z=\epsilon_y=0$  as:
\ba
M_s(a_{\rm out})=\int_{0}^{a_{\rm out}} 4\pi r^2\rho_s(r)dr=
2\pi\rho_s(a_{\rm out}) a_{\rm out}^3\propto a_{\rm out}^2.
\ea
For reference, we can take the density profile in Equation (\ref{eq:rho_gc}) with 
$\gamma=1.3$ for the Milky way's galactic center and 
set $M_{\rm BH}=4\times10^6M_\odot$ to get
\ba
\epsilon_{\rm sc}&\simeq&0.7\left(\frac{10\mbox{ AU}}{a}\right)^{3/2}
\left(\frac{a_{\rm out}}{0.1\mbox{ pc}}\right)^{3.2}
\left(\frac{m_1+m_2}{20M_\odot}\right)^{1/2}
(1-e_{\rm out}^2)^{3/2}.
\label{eq:omegas_explicit}
\ea
Using a general density profile changes the scaling to
$\epsilon_{\rm sc}\propto a_{\rm out}^{9/2-\gamma}$ and introduces a multiplicative 
factor of order unity
that depends on $\gamma$.


\begin{thebibliography}{}
\bibitem[Abbott et al.(2016a)]{2016PhRvL.116f1102A} Abbott, B.~P., Abbott, R., Abbott, T.~D., et al.\ 2016, Physical Review Letters, 116, 061102 
\bibitem[Abbott et al.(2016b)]{2016ApJ...833L...1A} Abbott, B.~P., Abbott, R., Abbott, T.~D., et al.\ 2016, \apjl, 833, L1 
\bibitem[Abbott et al.(2016)]{2016ApJ...818L..22A} Abbott, B.~P., Abbott, R., Abbott, T.~D., et al.\ 2016, \apjl, 818, L22 


\bibitem[Antognini et al.(2014)]{ASTA2014}
Antognini, J. M., Shappee, B. J., Thompson, T. A, \& Amaro-Seoane, P. 2014, 
\mnras, 439, 1

\bibitem[Antonini et al.(2010)]{2010ApJ...713...90A} Antonini, F., Faber, J., Gualandris, A., \& Merritt, D.\ 2010, \apj, 713, 90 
\bibitem[Antonini et al.(2011)]{2011ApJ...731..128A} Antonini, F., Lombardi, J.~C., Jr., \& Merritt, D.\ 2011, \apj, 731, 128 

\bibitem[Antonini et al.(2014)]{AMM2014}
Antonini, F., Murray, N., \& Mikkola, S. 2014, \apj, 781, 45

\bibitem[Antonini \& Perets(2012)]{2012ApJ...757...27A} Antonini, F., \& Perets, H.~B.\ 2012, \apj, 757, 27 


\bibitem[Antonini et al.(2015a)]{2015ApJ...812...72A} Antonini, F., Barausse, E., \& Silk, J.\ 2015, \apj, 812, 72 

\bibitem[Antonini et al.(2015b)]{2015ApJ...806L...8A} Antonini, F., Barausse, E., \& Silk, J.\ 2015, \apjl, 806, L8 

\bibitem[Antonini \& Rasio(2016)]{2016ApJ...831..187A} Antonini, F., \& Rasio, F.~A.\ 2016, \apj, 831, 187 

\bibitem[Antonini et al.(2017)]{ATH17}	
Antonini, F., Toonen, S., \& Hamers, A. S. 2017, arXiv:1703.06614


\bibitem[Bartko et al.(2010)]{2010ApJ...708..834B} 
Bartko, H., Martins, F., Trippe, S., et al.\ 2010, \apj, 708, 834 

\bibitem[Binney \& Tremaine(2008)]{BT08}
Binney, J., \& Tremaine, S. 2008, Galactic Dynamics: Second Edition (Princeton,
NJ: Princeton Univ. Press)

\bibitem[Blaes et al.(2002)]{blaes02}
Blaes, O., Lee, M. H., \& Socrates, A., 2002, \apj, 578, 775

\bibitem[Bortolas et al.(2017)]{2017arXiv170405850B} Bortolas, E., Mapelli, M., \& Spera, M.\ 2017, arXiv:1704.05850 

\bibitem[Bou\'e \& Fabrycky(2014)]{BF14}
Bou\'e, G., \& Fabrycky, D.C. 2014, \apj, 789, 110

\bibitem[Bradnick et al.(2017)]{BML17}
Bradnick, B., Mandel, I., \& Levin, Y. 2017, \mnras, 469, 2042

\bibitem[Carson et al.(2015)]{2015AJ....149..170C} Carson, D.~J., Barth, A.~J., Seth, A.~C., et al.\ 2015, \aj, 149, 170 

\bibitem[Chandrasekhar(1969)]{chandra69}
Chandrasekhar, S. 1969, The Silliman Foundation Lectures (New Haven, CT:
Yale Univ. Press)

\bibitem[Chatzopoulos et al.(2015)]{Chat15}
Chatzopoulos S., Fritz T. K., Gerhard O., Gillessen S. et al.
, 2015, \mnras, 447, 948

\bibitem[Conselice et al.(2005)]{2005ApJ...620..564C} Conselice, C.~J., Blackburne, J.~A., \& Papovich, C.\ 2005, \apj, 620, 564 

\bibitem[de Mink \& Belczynski(2015)]{2015ApJ...814...58D} de Mink, S.~E., \& Belczynski, K.\ 2015, \apj, 814, 58 

\bibitem[den Brok et al.(2014)]{2014MNRAS.445.2385D} den Brok, M., Peletier, R.~F., Seth, A., et al.\ 2014, \mnras, 445, 2385 

\bibitem[Dominik et al.(2012)]{2012ApJ...759...52D} Dominik, M., Belczynski, K., Fryer, C., et al.\ 2012, \apj, 759, 52 

\bibitem[Dosopoulou \& Antonini(2016)]{2016arXiv161106573D} Dosopoulou, F., \& Antonini, F.\ 2016, arXiv:1611.06573 

\bibitem[Dosopoulou \& Kalogera(2016a)]{2016ApJ...825...71D} Dosopoulou, F., \& Kalogera, V.\ 2016, \apj, 825, 71 

\bibitem[Dosopoulou \& Kalogera(2016b)]{2016ApJ...825...70D} Dosopoulou, F., \& Kalogera, V.\ 2016, \apj, 825, 70 

\bibitem[Duquennoy \& Mayor(1991)]{1991A&A...248..485D} Duquennoy, A., \& Mayor, M.\ 1991, \aap, 248, 485 

\bibitem[Duch{\^e}ne \& Kraus(2013)]{2013ARA&A..51..269D} Duch{\^e}ne, G., \& Kraus, A.\ 2013, \araa, 51, 269 

\bibitem[Fabrycky \& Tremaine(2007)]{FT07}
Fabrycky D. C. \& Tremaine, S. 2007, \apj, 669, 1298

\bibitem[Feldmeier-Krause et al.(2015)]{2015A&A...584A...2F} Feldmeier-Krause, A., Neumayer, N., Sch{\"o}del, R., et al.\ 2015, \aap, 584, A2 

\bibitem[Feldmeier-Krause et al.(2017)]{2017MNRAS.466.4040F} Feldmeier-Krause, A., Zhu, L., Neumayer, N., et al.\ 2017, \mnras, 466, 4040 

\bibitem[Figer et al.(2004)]{2004ApJ...601..319F} Figer, D.~F., Rich, R.~M., Kim, S.~S., Morris, M., \& Serabyn, E.\ 2004, \apj, 601, 319 

\bibitem[Forbes et al.(2008)]{2008MNRAS.389.1924F} Forbes, D.~A., Lasky, P., Graham, A.~W., \& Spitler, L.\ 2008, \mnras, 389, 1924 

\bibitem[Ford et al.(2000)]{2000ApJ...535..385F} Ford, E.~B., Kozinsky, B., \& Rasio, F.~A.\ 2000, \apj, 535, 385 

 
 \bibitem[Fritz et al.(2016)]{2016ApJ...821...44F} Fritz, T.~K., Chatzopoulos, S., Gerhard, O., et al.\ 2016, \apj, 821, 44 

 \bibitem[Fryer \& Kalogera(2001)]{2001ApJ...554..548F} Fryer, C.~L., \& Kalogera, V.\ 2001, \apj, 554, 548 

\bibitem[Georgiev \& B{\"o}ker(2014)]{2014MNRAS.441.3570G} Georgiev, I.~Y., \& B{\"o}ker, T.\ 2014, \mnras, 441, 3570 
\bibitem[Georgiev et al.(2016)]{2016MNRAS.457.2122G} Georgiev, I.~Y., B{\"o}ker, T., Leigh, N., L{\"u}tzgendorf, N., \& Neumayer, N.\ 2016, \mnras, 457, 2122 

\bibitem[Ghez et al.(2008)]{2008ApJ...689.1044G} Ghez, A.~M., Salim, S., Weinberg, N.~N., et al.\ 2008, \apj, 689, 1044-1062 
\bibitem[Gillessen et al.(2009)]{2009ApJ...692.1075G} Gillessen, S., Eisenhauer, F., Trippe, S., et al.\ 2009, \apj, 692, 1075 

\bibitem[Hamers et al.(2015)]{hamers15}
Hamers, A. S., Perets, H. B., Antonini, F., \& Portegies Zwart, S. F. 2015,
\mnras, 449, 4221

\bibitem[Hamers \& Lai(2017)]{HL17}
Hamers, H. S., \& Lai, D. 2017, arXiv:1705.02334

\bibitem[Heggie(1975)]{H75}
Heggie, D. C. 1975, \mnras, 173, 729

\bibitem[Hobbs et al.(2005)]{2005MNRAS.360..974H} Hobbs, G., Lorimer, D.~R., Lyne, A.~G., \& Kramer, M.\ 2005, \mnras, 360, 974 

\bibitem[Hopman \& Alexander(2006)]{2006ApJ...645.1152H} Hopman, C., \& Alexander, T.\ 2006, \apj, 645, 1152 

\bibitem[Holman et al. (1997)]{holman97}
Holman, M., Touma, J., \& Tremaine, S. 1997, Nature, 386, 254

\bibitem[Hurley et al.(2002)]{2002MNRAS.329..897H} Hurley, J.~R., Tout, C.~A., \& Pols, O.~R.\ 2002, \mnras, 329, 897 

\bibitem[Ivanov et al.(2005)]{ivanov05}
Ivanov, P. B., Polnarev, A. G., \& Saha, P. 2005, \mnras, 358, 1361

\bibitem[Kalogera et al.(2001)]{2001ApJ...556..340K} Kalogera, V., Narayan, R., Spergel, D.~N., \& Taylor, J.~H.\ 2001, \apj, 556, 340 


\bibitem[Kalogera \& Webbink(1998)]{1998ApJ...493..351K} Kalogera, V., \& Webbink, R.~F.\ 1998, \apj, 493, 351 

 \bibitem[Katz et al.(2011)]{katz11} 
 Katz, B., Dong, S., \& Malhotra, R.\ 2011, Physical Review Letters, 107, 181101 

\bibitem[Katz \& Dong(2012)]{KD12}
Katz, B. \& Dong, S. 2012, arXiv:1211.4584

\bibitem[Kobulnicky et al.(2014)]{2014ApJS..213...34K} 
Kobulnicky, H.~A., Kiminki, D.~C., Lundquist, M.~J., et al.\ 2014, \apjs, 213, 34 

\bibitem[Kocsis \& Tremaine(2011)]{KT11}
Kocsis B., \& Tremaine S., 2011, \mnras, 412, 187

\bibitem[Kopparapu et al.(2008)]{2008ApJ...675.1459K} Kopparapu, R.~K., Hanna, C., Kalogera, V., et al.\ 2008, \apj, 675, 1459-1467 

\bibitem[Kozai(1962)]{1962AJ.....67..591K} Kozai, Y.\ 1962, \aj, 67, 591 

\bibitem[Kroupa(2002)]{2002Sci...295...82K} Kroupa, P.\ 2002, Science, 295, 82 

\bibitem[Lauer et al.(1998)]{1998AJ....116.2263L} Lauer, T.~R., Faber, S.~M., Ajhar, E.~A., Grillmair, C.~J., \& Scowen, P.~A.\ 1998, \aj, 116, 2263 

\bibitem[Lef{\`e}vre et al.(2009)]{2009A&A...507.1141L} Lef{\`e}vre, L., Marchenko, S.~V., Moffat, A.~F.~J., \& Acker, A.\ 2009, \aap, 507, 1141 

\bibitem[Li et al.(2014)]{li14}
Li, G., Naoz, S., Holman, M., \& Loeb, A. 2014, \apj, 791, 86

\bibitem[Lichtenberg \& Lieberman(1983)]{LL83}
Lichtenberg, A. J., \& Lieberman, M. A. 1983, 
Regular and Chaotic Dynamics (New York: Springer)

\bibitem[Lidov(1962)]{1962P&SS....9..719L} Lidov, M.~L.\ 1962, \planss, 9, 719 

\bibitem[Liu et al.(2015)]{liu15} 
Liu, B., Mu{\~n}oz, D.~J., \& Lai, D.\ 2015, \mnras, 447, 747 

\bibitem[Liu et al.(2017)]{2017MNRAS.466.3376L} Liu, B., Wang, Y.-H., \& Yuan, Y.-F.\ 2017, \mnras, 466, 3376

\bibitem[Lithwick \& Naoz(2011)]{LN11}
Lithwick, Y., \& Naoz, S. 2011, \apj, 742, 94

\bibitem[L{\"o}ckmann et al.(2010)]{2010MNRAS.402..519L} L{\"o}ckmann, U., Baumgardt, H., \& Kroupa, P.\ 2010, \mnras, 402, 519 

\bibitem[Lu et al.(2013)]{2013ApJ...764..155L} Lu, J.~R., Do, T., Ghez, A.~M., et al.\ 2013, \apj, 764, 155 

\bibitem[Magorrian \& Tremaine(1999)]{MT99}
Magorrian, J., \& Tremaine, S. 1999, \mnras, 309, 447

\bibitem[Merritt et al.(2011)]{merritt2011}
Merritt, D., Alexander, T., Mikkola, S., \& Will, C. M. 2011, PhRvD, 84,
044024

\bibitem[Merritt \& Vasiliev(2011)]{MV11}
Merritt, D., \& Vasiliev, E. 2011, \apj, 726, 61

\bibitem[Merritt(2013)]{2013degn.book.....M} Merritt, D.\ 2013, 
Dynamics and Evolution of Galactic Nuclei,
Princeton, NJ, Princeton University Press, 2013

\bibitem[Milankovitch(1939)]{M39}
Milankovitch, M., 1939, Bull. Serb. Acad. Math. Nat. A 6, 1

\bibitem[Misgeld \& Hilker(2011)]{2011MNRAS.414.3699M} Misgeld, I., \& Hilker, M.\ 2011, \mnras, 414, 3699 

\bibitem[Naoz et al.(2013)]{naoz13} 
Naoz, S., Farr, W. M., Lithwick, Y., Rasio, F. A., \& 
Teyssandier, J. 2013, \mnras, 431, 2155

\bibitem[Naoz(2016)]{naoz16}
Naoz S., 2016, ARA\&A, 54, 441

\bibitem[Neumayer et al.(2011)]{2011MNRAS.413.1875N} Neumayer, N., Walcher, C.~J., Andersen, D., et al.\ 2011, \mnras, 413, 1875 

\bibitem[Neumayer \& Walcher(2012)]{2012AdAst2012E..15N} Neumayer, N., \& Walcher, C.~J.\ 2012, Advances in Astronomy, 2012, 709038 

\bibitem[Pejcha et al.(2013)]{PAST13}
Pejcha, O., Antognini, J. M., Shappee, B. J., \& Thompson T. A., 2013, MNRAS,
435, 943

\bibitem[Peters(1964)]{1964PhRv..136.1224P} Peters, P.~C.\ 1964, Physical Review, 136, 1224 

\bibitem[Petrovich(2015)]{2015ApJ...805...75P} 
Petrovich, C.\ 2015, \apj, 799, 27

\bibitem[Petrovich \& Mu\~noz(2017)]{PM17}
Petrovich, C. \& Mu\~noz, D. J. 2017, \apj, 834, 116

\bibitem[Pfuhl et al.(2014)]{2014ApJ...782..101P} Pfuhl, O., Alexander, T., Gillessen, S., et al.\ 2014, \apj, 782, 101 

\bibitem[Pfuhl et al.(2011)]{2011ApJ...741..108P} Pfuhl, O., Fritz, T.~K., Zilka, M., et al.\ 2011, \apj, 741, 108 

\bibitem[Prodan et al.(2015)]{2015ApJ...799..118P} Prodan, S., Antonini, F., \& Perets, H.~B.\ 2015, \apj, 799, 118 

\bibitem[Rauch \& Tremaine(1996)]{1996NewA....1..149R} Rauch, K.~P., \& Tremaine, S.\ 1996, New Astronomy, 1, 149 

\bibitem[Rodriguez et al.(2016)]{2016PhRvD..93h4029R} Rodriguez, C.~L., Chatterjee, S., \& Rasio, F.~A.\ 2016, \prd, 93, 084029 

\bibitem[Rossa et al.(2006)]{2006AJ....132.1074R} Rossa, J., van der Marel, R.~P., B{\"o}ker, T., et al.\ 2006, \aj, 132, 1074 

\bibitem[Sambhus \& Sridhar(2000)]{SS00}
Sambhus, N., \& Sridhar, S. 2000, \apj, 542, 143

\bibitem[Sana et al.(2012)]{2012Sci...337..444S} Sana, H., de Mink, S.~E., de Koter, A., et al.\ 2012, Science, 337, 444 

\bibitem[Sch{\"o}del et al.(2014)]{2014A&A...566A..47S} Sch{\"o}del, R., Feldmeier, A., Kunneriath, D., et al.\ 2014, \aap, 566, A47 

\bibitem[Sch{\"o}del et al.(2017)]{2017arXiv170103817S} Sch{\"o}del, R., Gallego-Cano, E., Dong, H., et al.\ 2017, arXiv:1701.03817 


\bibitem[Seth et al.(2008)]{2008ApJ...678..116S} Seth, A., Ag{\"u}eros, M., Lee, D., \& Basu-Zych, A.\ 2008, \apj, 678, 116-130


\bibitem[Spitzer(1987)]{1987degc.book.....S} Spitzer, L.\ 1987, 
Princeton, NJ, Princeton University Press, 1987, 191 p.,  

\bibitem[Stephan et al.(2016)]{2016MNRAS.460.3494S} 
Stephan, A.~P., Naoz, S., Ghez, A.~M., et al.\ 2016, \mnras, 460, 3494  

\bibitem[Sridhar \& Touma(1999)]{ST99}
Sridhar, S., \& Touma, J. 1999, \mnras, 303, 483

\bibitem[Tremaine et al.(2009)]{TTN09}
Tremaine S., Touma J., \& Namouni F., 2009, AJ, 137, 3706

\bibitem[Tremaine \& Yavetz(2014)]{TY14}
Tremaine, S. \& Yavetz, T. 2014, arXiv:1309.5244

\bibitem[Turner et al.(2012)]{2012ApJS..203....5T} Turner, M.~L., C{\^o}t{\'e}, P., Ferrarese, L., et al.\ 2012, \apjs, 203, 5 

\bibitem[VanLandingham et al.(2016)]{2016ApJ...828...77V} VanLandingham, J.~H., Miller, M.~C., Hamilton, D.~P., \& Richardson, D.~C.\ 2016, \apj, 828, 77 

\bibitem[White et al.(1987)]{white87}
White, R. E., \& Shawl, S. J. 1987, \apj, 317, 246

\bibitem[Yelda et al.(2014)]{2014ApJ...783..131Y} Yelda, S., Ghez, A.~M., Lu, J.~R., et al.\ 2014, \apj, 783, 131 


\end{thebibliography}

\end{document}